\documentclass[fleqn,usenatbib]{mnras}
\usepackage{newtxtext,newtxmath}
\usepackage[T1]{fontenc}

\usepackage{graphicx}	
\usepackage{amsmath}	
\usepackage{comment}
\usepackage{xcolor}
\usepackage{subcaption}

\newcommand{\pfrac}[2]{\left( \frac{#1}{#2} \right)}

\newcommand{\rp}{r_{\rm p}}
\newcommand{\rpmin}{r_{\rm p,min}}

\newcommand{\rtidal}{r_{\rm t}}
\newcommand{\rh}{r_{\rm h}}
\newcommand{\rtidalb}{r_{\rm tidal,b}}

\newcommand{\MBH}{M_{\bullet}}
\newcommand{\rhod}{\rho_{\rm d}}
\newcommand{\Sigmad}{\Sigma_{\rm d}}
\newcommand{\kb}{k_{\rm B}}

\title{Period Evolution of Repeating Transients in Galactic Nuclei}

\author[Linial \& Quataert]{
Itai Linial,$^{1,2}$\thanks{E-mail: itai.linial@mail.huji.ac.il}
Eliot Quataert,$^{3}$
\\
$^{1}$School of Natural Sciences, Institute for Advanced Study, Princeton, NJ 08540, USA\\
$^{2}$Columbia Astrophysics Laboratory, Columbia University, New York, NY 10027, USA\\
$^{3}$Department of Astrophysical Sciences, Princeton University, Peyton Hall, Princeton, NJ 08540, USA\\
}

\date{Accepted XXX. Received YYY; in original form ZZZ}

\pubyear{2023}

\begin{document}
\label{firstpage}
\pagerange{\pageref{firstpage}--\pageref{lastpage}}
\maketitle

\begin{abstract}
Wide-field survery have recently detected recurring optical and X-ray sources near galactic nuclei, with period spanning hours to years. These phenomena could result from repeated partial tidal disruptions of stars by supermassive black holes (SMBHs) or by interaction between star and SMBH-accretion discs. We study the physical processes that produce period changes in such sources, highlighting the key role of the interaction between the orbiting star and the accretion disc. We focus on ASASSN-14ko - a repeatedly flaring optical source with a mean period $P_0 = 115 \, \rm d$ and a detected period decay $\dot{P} = -2.6\times 10^{-3}$ (Payne et al. 2022). We argue that the system's $\dot{P}$ is most compatible with true orbital decay produced by hydrodynamical drag as a star passes through the accretion disc on an inclined orbit, twice per orbit. The star is likely a sun-like star whose envelope is somewhat inflated, possibly due to tidal heating. Star-disc interaction inevitably leads to drag-induced stripping of mass from the star, which may be the dominant component in powering the observed flares. We discuss ASASSN-14ko's possible formation history and observational tests of our interpretation of the measured $\dot P$. Our results imply that partial tidal disruption events manifesting as repeating nuclear transients cannot be modeled without accounting for the cumulative impact of tidal heating over many orbits. We discuss the implications of our results for other repeating transients, and predict that the recurrence time of Quasi-Periodic Eruptions is expected to decay at a rate of order $|\dot{P}| \approx 10^{-6}-10^{-5}$.
\end{abstract}




\section{Introduction}
Wide-field optical and X-ray surveys have recently detected repeating flares in centers of galaxies, with recurrence times ranging from a few to several hours \citep[Quasi-Periodic Eruptions, QPEs, e.g.,][]{Miniutti_2019,Giustini_2019,Arcodia_2021,Chakraborty_2021}, up to years-decade timescale \citep{Payne_2021,Liu_obs_2023,Wevers_2022b,Malyali_2023}. These recent discoveries likely represent a subset of a broader class of periodic nuclear transients (PNTs), extending to longer and shorter recurrence timescales than those identified thus far. As transient surveys continue to monitor flaring sources over longer baselines, new temporal regimes are expected to be unveiled.

While the physical origin of these systems remains unknown, their recurring nature,
 association with galactic nuclei, and active galactic nuclei luminosities suggest a binary origin, with a stellar object orbiting the galaxy's supermassive black hole (SMBH).   Partial  disruption at every pericenter passage or interaction between the orbiting object and an accretion flow can then power the bright X-ray/optical flares \citep{Nayashkin_2004,King_2020,Zhao_2021,Xian_2021, King_2022,Krolik_Linial_2022,Lu_Quataert_2023,Linial_Sari_2023,Metzger_2021,Sukova_2021,Payne_2021,Payne_2022, Linial_Metzger_23,Tagawa_23,Franchini_23}. 
 
One particularly promising partial/repeating TDE candidate is ASASSN-14ko - an optically detected periodically flaring nuclear transient located at the center of the galaxy ESO 253-G003, with a mean recurrence interval of $P_0 \approx 115 \, \rm d$ \citep{Payne_2021,Payne_2022}. The regular timing of this system's flares is suggestive of a binary origin, consistent with its interpretation as a repeating partial TDE \citep[e.g.,][]{Payne_2021,Liu2023}. However, over nearly a decade of observations, the system's period has systematically decreased at a rate $\dot{P}\approx -0.0026 \approx -(7 \, \textrm{hr})/P_0$. Such orbital decay is much larger than predicted by gravitational-wave inspiral of a companion \citep[e.g.,][]{Payne_2021,Cufari_2022}. Another recent related system is the X-ray source eRASSt J045650.3-203750 \citep{Liu_obs_2023}, flaring roughly every $\sim 223 \, \rm d$. Its recurrence time appears to have evolved by roughly $10\%$ over the course of a few flares (Liu, private communication). Lastly, recently identified QPEs appear to undergo secular period evolution on year-decade timescales \citep[e.g.,][]{Miniutti_23a,Miniutti_23}.

The main focus of this paper will be on the importance of interaction between the binary companion and the SMBH's accretion flow in driving period evolution in the timing of the flares of repeating nuclear transients. Similar themes have been explored by several authors, including \cite{Syer_1991,Vokrouhlicky_1993,Vokrouhlicky_1998a,Vokrouhlicky_1998b} and more recently \cite{Macleod_Lin_2020},\cite{Generozov_23} and \cite{Wang_23}. We also consider a variety of other sources of period evolution in order to provide a comprehensive analysis that we hope will be useful for interpreting the emerging population of PNTs.  We specifically focus on  ASASSN-14ko as a concrete example due to its superb timing characterization, but much of our analysis will be general and can readily applied to other systems.  We shall show that for the specific case of ASASSN-14ko, the period evolution implies strong constraints on the nature of the orbiting star and the system's interpretation as a repeating partial TDE.

The paper is organized as follows: We review the relevant observations of ASASSN-14ko and their implications regarding the orbit in \S \ref{sec:14ko_properties}. We then consider the interaction between a star on a highly inclined orbit and an accretion disc in \S \ref{sec:star_disc}, and then apply these results to ASASSN-14ko and discuss their implications in \S \ref{sec:14ko_Pdot_application}. Several alternative sources of period evolution are reviewed in \S \ref{sec:other}, where we argue that they are incompatible with observations of ASASSN-14ko (though they in principle could be applicable to other systems).  We briefly discuss the connection between ASASSN-14ko's formation channel and its period evolution in \S \ref{sec:formation}, before concluding and summarizing the paper in \S \ref{sec:discussion}.

\section{Orbital Properties and Constraints on ASASSN-14ko} \label{sec:14ko_properties}

ASASSN-14ko is a periodically flaring active galactic nucleus (AGN) in the galaxy ESO 253-G003.  \citet{Tucker2021} show that the host galaxy is an onging merger, with two AGNs in separate galactic nuclei separated by about 1.4 kpc; the periodic flares are associated with the north-eastern nucleus.   ASASSN-14ko's $\sim 20$ currently observed flares reveal a mean recurrence time of $P_0=115.6 \pm 2.2 \, \rm days$, with an observed period evolution of $\dot{P} = -0.0026 \pm 0.0006$ \citep{Payne_2021,Payne_2022}. The system's nearly perfect periodicity suggest an underlying clock set by an orbiting low-mass companion, as argued in the discovery papers.  Given the estimates of the SMBH mass, $\MBH \approx 10^7-10^8 \, \rm M_\odot$ \citep{Payne_2021}, the orbit's semi-major axis is
\begin{equation} \label{eq:a_from_P}
    a = (G\MBH (P/(2\pi))^2)^{1/3} \approx 100 \; \mathcal{P}_{115}^{2/3} M_{\bullet,7}^{1/3} \; \rm au \,,
\end{equation}
where $\mathcal{P}_{115} = P_0/(115 \, \rm days)$, $M_{\bullet,7} = \MBH/10^7 \, \rm M_\odot$.

Since the companion survives multiple orbits, its pericenter distance, $\rp$, is limited by the tidal radius
\begin{equation}
    \rtidal = R_\star \pfrac{\MBH}{m_\star}^{1/3} \approx 1 \; R_1 m_1^{-1/3} M_{\bullet,7}^{1/3} \; \rm au \,,
\end{equation}
where $m_1 = m_\star/{\rm M_\odot}$ and $R_1 = R_\star/{\rm R_\odot}$. Note that as $\rp$ approaches a few times $\rtidal$, tides may cause the star to puff up and overflow its Roche lobe earlier than it would have had the stellar structure remained unaffected by tidal heating. Substantial heating takes place when $\dot{E}_{\rm tides} = E_{\rm tides}/P \gtrsim L_\star$, where $E_{\rm tides}$ is the energy deposited in stellar tides at pericenter passage and $L_\star$ is the star's equilibrium luminosity. This condition implies a critical peri-center separation, $r_{\rm MT}$, marking the onset of mass transfer initiated by tidal heating
\begin{equation} \label{eq:r_MT}
    r_{\rm MT} \approx 4.6 \; \rtidal \, \mathcal{P}_{115}^{-0.08} \,,
\end{equation}
where the weak dependence on $P$ is a consequence of the very strong dependence of $E_{\rm tides}$ on $\rp/\rtidal$ \citep[e.g.,][]{Press_Teukolsky_1977,Lee_Ostriker_1986}. The derivation of the above condition and further discussion are given in Appendix \ref{appendix:tides}. As the above criterion is met, the star becomes more susceptible to tidal heating, leading to runaway increase in its radius, up to the onset of mass transfer, occurring at $\rp \approx 2 R_\star (\MBH/m_\star)^{1/3}$ once it overflows it Roche lobe at pericenter\footnote{In the commonly used notation of $\beta \equiv \rtidal/\rp$, a partial TDE would occur when $\beta = \beta_{\rm c} \approx 0.5$ \citep[e.g.,][]{Ryu2020}. Here we account for the long-term tidal heating of the star incurred over multiple pericenter passages, initiating mass loss at a somewhat larger $\rp$ compared to an undisturbed star placed on parabolic orbit passing close to the SMBH.}. 
In what follows, we will generally consider both $\rp \approx r_{\rm MT}$ as well as $\rp \gtrsim {\rm few} \, \times \, r_{\rm MT}$, and will therefore allow for an order-unity uncertainty regarding the minimal value of $\rp$, probably a few times $\rtidal$.

The constraint on the pericenter radius $\rp \gtrsim \rtidal$, corresponds to an upper limit on the orbital eccentricity of
\begin{equation}
    1-e  = \frac{\rp}{a} \gtrsim \frac{\rtidal}{a} = 0.01 \; R_1 m_1^{-1/3} \mathcal{P}_{115}^{-2/3} \,.
\end{equation}
Finally, since $\rtidal \leq \rp < a$, the star's mean density must exceed
\begin{equation}
    \bar{\rho}_\star \equiv \frac{m_\star}{4\pi R_\star^3/3} > \frac{24 \pi}{G P^2} \approx 10^{-5} \; \mathcal{P}_{115}^{-2} \; \rm g \; cm^{-3} \,,
\end{equation}
or equivalently, the companion's radius is smaller than
\begin{equation}
    R_\star < \frac{1}{2} (Gm_\star (P/2\pi)^2)^{1/3} \approx 50 \; m_1^{1/3} \mathcal{P}_{115}^{2/3} \; \rm R_\odot \,,
\end{equation}
ruling out stars with an extended envelope, $R_1 \gtrsim 50 \, m_1^{1/3}$.

For highly compact companions, with mean density exceeding $\bar{\rho}_\star > c^6 / (36 \pi G^3 \MBH^2) \approx 55 \; M_{\bullet,7}^{-2} \; \rm g \, cm^{-3}$, tidal gravity is always negligible, since $\rtidal < r_{\rm ISCO} \approx 6G\MBH/c^2 \approx 0.6 \; M_{\bullet,7} \; \rm au$. In this case, the pericenter distance must be greater than $r_{\rm ISCO}$ to avoid direct capture by the SMBH.

\subsection{General arguments regarding the emission} \label{sec:general_emission}
With a peak flare luminosity of $L_{\rm p} \approx 2\times 10^{44} \, \rm erg \, s^{-1}$ and a duration of $\Delta t \approx 10 \, \rm d$, the total radiated energy per flare is of order $E_{\rm rad} \approx L_{\rm p} \Delta t \approx 2 \times 10^{50} \, \rm erg$. The fitted black body temperature is roughly $T_{\rm p} \approx 10^4 \, \rm K$, and the corresponding radius of the emitting region is approximately $r_e \approx 10^{15} \; {\rm cm} \approx 70 \; \rm au$.

This amount of radiated energy can be provided by a mass per-flare 
\begin{equation} \label{eq:dM_rad}
    \delta m_{\rm rad} \approx E_{\rm rad}/(\eta c^2) \approx 10^{-3} \eta_{0.1}^{-1} \, \rm M_\odot \,,
\end{equation} 
where $\eta$ is the conversion efficiency from rest mass to radiated optical energy, and $\eta_{0.1} = \eta/0.1$. Accounting for the $N_{\rm obs} \approx 20$ observed flares since the beginning of the system's monitoring, the required mass budget is at least $\delta m_{\rm tot} > 0.02 \eta_{0.1}^{-1} \; \rm M_\odot$.

An efficiency of $\eta \approx 0.1$ can indeed be achieved if the observed flares are produced through accretion onto the SMBH. However, the accretion luminosity will likely be in the UV and X-rays, which may then be reprocessed to optical emission through an extended distribution of gas. Na\"ively, accretion is delayed by the viscous time at the orbit's pericenter
\begin{equation}
    \label{eq:tvisc}
    t_{\rm visc} \approx \sqrt{ \frac{\rp^3}{G\MBH} } \alpha^{-1} \pfrac{\rp}{h}^2 \approx 1.8 \; \pfrac{\rp}{1 \, \rm au}^{3/2} M_{\bullet,7}^{-1/2} \alpha^{-1} \pfrac{\rp/h}{10}^{2} \rm day \,,
\end{equation}
where $h$ is the accretion disc's scale-height.

\textit{Swift} and \textit{NICER} X-ray observations of ASASSN-14ko show quiescent emission between flares at a level of roughly $L_{\rm Q} \approx 10^{43} \, \rm erg \, s^{-1}$ \citep{Payne_2021,Payne_2022}, likely originating from an AGN disc accreting at an Eddington ratio of roughly
\begin{equation}
    \dot{m}/\dot{m}_{\rm Edd} \approx 8\times 10^{-3} \; \pfrac{L_{\rm Q}}{10^{43} \; \rm erg \, s^{-1}} M_{\bullet,7}^{-1} \,.
    \label{eq:mdot_q}
\end{equation}
The total energy emitted in X-rays during quiescence over a full period is comparable to the energy emitted in a single flare, $E_{\rm rad,Q} = L_{\rm Q} P_0 \approx 10^{50} \; {\rm erg} \lesssim E_{\rm rad}$. We also note that at the accretion rate in equation \ref{eq:mdot_q}, standard radiation dominated disc models suggest that the disc should be quite thin, with $h/\rp \sim 10^{-2}$ (eq. \ref{eq:h_rad_disc} discussed below) and thus $t_{\rm visc} \sim 200 \alpha^{-1}$ days, longer than the period of the flares and much longer than the flare duration. This is a challenge to powering the observed emission directly by accretion. It is possible, however, that the circularization of material stripped from the companion at pericenter or during star-disc interactions leads to much more rapid redistribution of angular momentum, as in the stream-stream shocks that occur in TDEs \citep{Lu2020}.

Tidally stripped material attains a spread in specific orbital energy, of order
\begin{equation}
    \Delta E \approx \frac{G m_\star}{R_\star} \pfrac{\MBH}{m_\star}^{1/3} \,.
\end{equation}

If $\Delta E \gg G\MBH/2a$, a significant fraction of the debris is more tightly bound to the SMBH than the star, falling back on a timescale
\begin{equation}
    t_{\rm fb} \approx  \sqrt{ \frac{\rp^3}{G\MBH} } \pfrac{\MBH}{m_\star}^{1/2} \approx 60 \; \pfrac{\rp}{1 \, \rm au}^{3/2} m_1^{-1/2} \; \rm day \,,
\end{equation}
with $t_{\rm fb} \ll P$, completeing several orbits before the star's next pericenter passage. Within this time frame, gas may circularize and form a disc of size $\sim \rp$ (with which the star may interact in its subsequent passage). If in addition $t_{\rm visc} \ll P$, most of the circularized gas is accreted before the star's next pericenter visit.

On the contrary, if $\Delta E \ll G\MBH/2a$, essentially all of the stripped mass remains bound to the SMBH, with an orbital period similar to that of the star, $P$. In this regime, circularization is not necessarily achieved fast enough, and the nature of the coupling between the star and the stripped matter may be quite different from that previously described.  

Note that in the case of ASASSN-14ko, $\Delta E \approx E_{\rm orb}$ (or equivalently, $t_{\rm fb} \approx P_0$), hence it is not entirely clear which is the appropriate regime. This is due to the fact that $\rtidal/a \approx (m_\star/\MBH)^{1/3}$ in this system. Understanding circularization of tidally stripped debris in this regime would be valuable as it is likely important for understanding whether accretion can power the flares (see discussion after eq. \ref{eq:tvisc}) and whether stripped gas at $\sim 10^{15}$ cm covers a sufficient solid angle to explain the reprocessed optical radiation.

\section{Star-disc Interaction} \label{sec:star_disc}
Motivated by the orbiting companion interpretation of PNTs including ASASSN-14ko, we consider the interaction of an orbiting star and an accretion disc feeding the SMBH. The star is assumed to orbit the SMBH on a highly inclined, highly eccentric orbit, with $\iota \sim \mathcal{O}(1)$ and $1-e\ll 1$. {A similar drag model was considered by \cite{Macleod_Lin_2020} to capture the effects of star-disc interaction on loss-cone dynamics in active galactic nuclei.}

We assume an optically thick, geometrically thin, radiatively efficient disc, characterized by a constant $\alpha$-parameter and a constant mass accretion rate $\dot{m}$. As the star passes through the disc midplane, it is subject to hydrodynamical drag force whose magnitude is approximately
\begin{equation}
    |\mathbf{f}_{\rm drag}| \approx \frac{1}{2}\rhod v_{\rm rel}^2 \pi R_{\rm drag}^2 \,,
\end{equation}
where $p_{\rm ram} \approx \frac{1}{2} \rhod v_{\rm rel}^2$ is the ram pressure exerted by gas of density $\rhod$ on the companion, moving at velocity $v_{\rm rel}$ relative to the impacted gas. The ram pressure acts upon an effective cross section, $\pi R_{\rm drag}^2$, where $R_{\rm drag} \approx \max{ \{ R_\star,Gm_\star/v_{\rm rel}^2 \} }$, namely the greater of the companion's physical radius and its Bondi-Holye radius. For a main-sequence stellar companion, and not a compact object, its effective radius will typically be $R_\star$, provided that $v_{\rm rel} \gtrsim v_{\rm esc}$ where $v_{\rm esc}$ is its surface escape velocity. {For a sun-like star, this condition is satisfied throughout the entire orbit for $a \leq R_\star (\MBH/m_\star) \approx 0.2 \, {\rm pc} \; R_1 M_{\bullet,7} m_1^{-1}$}. We address the complementary regime in appendix \ref{appendix:graviational_focusing}.

We model the effect of the star's passage through the disc as an impulsive change to its orbital velocity resulting from the momentum imparted to the intercepted disc material, namely
\begin{equation} \label{eq:deltaV_drag}
    \mathbf{v}' \approx \mathbf{v}\left( 1 - \frac{W_{\rm drag}}{2 E_k}\right) \approx \mathbf{v}\left( 1- \frac{\pi \Sigmad R_\star^2}{2 m_\star} \right) \,,
\end{equation}
where $\mathbf{v}$ and $\mathbf{v}'$ are the velocities before and after disc passage, $W_{\rm drag}$ is the work done by the drag force during disc passage, $E_k$ is the star's kinetic energy at the time of collision, and $\Sigmad = 2 \rhod h$ is the disc surface density at the collision site. Here we assumed that the inclination angle, $\iota$, is close to $\pi/2$, such that the distance traversed by the star through the disc in each collision is of order $2h$ and that $v_{\rm rel} \approx v_k$, comparable to the Keplerian velocity \footnote{{Star-disc interaction may dissipate the star's inclination, bringing it into alignment with the plane of the disc, in which case our assumption $\iota \sim \mathcal{O}(1)$ becomes invalid. However, the dissipation of orbital eccentricity is expected to occur on a similar timescale as the orbital alignment, as both involve a significant change in the orbital angular momentum due to drag induced torques.  For the case of ASASSN-14ko, since the orbit is currently highly eccentric, it is unlikely that the initial inclination has decayed to $\iota \ll 1$. If alternatively the star formed in the disc, the star would not have a large eccentricity, since the timescale to dissipate eccentricity is much less than the stellar lifetime.}}. We have assumed that the drag force is parallel to the star's velocity, i.e., the star is slightly decelerated by the headwind it experiences as it ploughs through the disc, neglecting the (mostly azimuthal) flow of gas. {\cite{Macleod_Lin_2020} used a somewhat more accurate drag prescription, accounting for the relative velocity between the star and the impacted disc material, and considered the entire range of orbital inclinations. Despite our somewhat more approximate treatment, we arrive at overall similar conclusions regarding the impact of accretion disc gas drag on the orbital evolution of stars on highly eccentric orbits.}

The location within the disc at which collisions occur gradually evolves due to relativistic apsidal precession, completing a full cycle over a period
\begin{equation} \label{eq:T_prec_ap}
    T_{\rm prec, ap} \approx P_0 \frac{a (1-e^2)}{3 R_g} = P_0 \pfrac{\rp}{R_g} \pfrac{1+e}{3} \,,
\end{equation}
and hence the experienced drag is expected to be modulated over the precession timescale.

Insofar as the fractional change to the star's linear momentum is small, namely, $\pi \Sigmad R_\star^2/m_\star \ll 1$ (which is very well satisfied for our typical parameters), the orbital energy is gradually dissipated by the interaction with the disc over many orbits. Averaged over the apsidal precession cycle, the star's orbital period decays as
\begin{equation} \label{eq:Pdot_drag_avg}
    \left< \dot{P}_{\rm orb} \right> \approx -\frac{3}{2} \frac{\left< W_{\rm drag} \right>}{|E_{\rm orb}|} \approx - \frac{3\pi R_\star^2 a}{m_\star} \left< \Sigmad(r) / r \right> \,,
\end{equation}
where $\left< W_{\rm drag} \right>$ is the work done by the drag force, per-orbit, averaged appropriately over a full precession cycle, and $E_{\rm orb} = -G\MBH m_\star / 2a$ is the star's (negative) orbital energy. 

If the extent of the disc is greater than $r_{\rm max} = a(1+e)$,
star disc-collisions will occur twice per orbit , with one inner collision occurring at a distance $\rp \lesssim r_1 \lesssim (1+e) \rp$, and an additional outer collision at $r_2$, where $r_1 \leq r_2 \leq r_{\rm max}$. If $\Sigmad(r)/r$ decreases with $r$ for $r > \rp$ {(which is typically the regime of interest, see for example Fig. \ref{fig:ThinDisc})}, collisions at $r\approx \rp$ dominate the star's orbital decay, and equation \ref{eq:Pdot_drag_avg} can then be rewritten as
\begin{equation} \label{eq:Pdot_avg_rp}
    \left< \dot{P}_{\rm orb} \right> \approx - \frac{3\pi \Sigmad(\rp) R_\star^2 a}{m_\star \rp} \,,
\end{equation}
The above expression can be easily understood through the following scaling argument - the fractional change in the star's momentum in a single collision is roughly $\pi R_{\star}^2 \Sigmad/m_\star$. If collisions at around $\rp$ dominate, the fractional change in orbital energy and period is enhanced by a factor $(a/\rp)$, corresponding to the ratio of lever arms, yielding the scaling of equation \ref{eq:Pdot_avg_rp}.

\begin{figure}      \centering
     \begin{subfigure}[b]{0.5\textwidth}
         \centering
         \includegraphics[width=\linewidth]{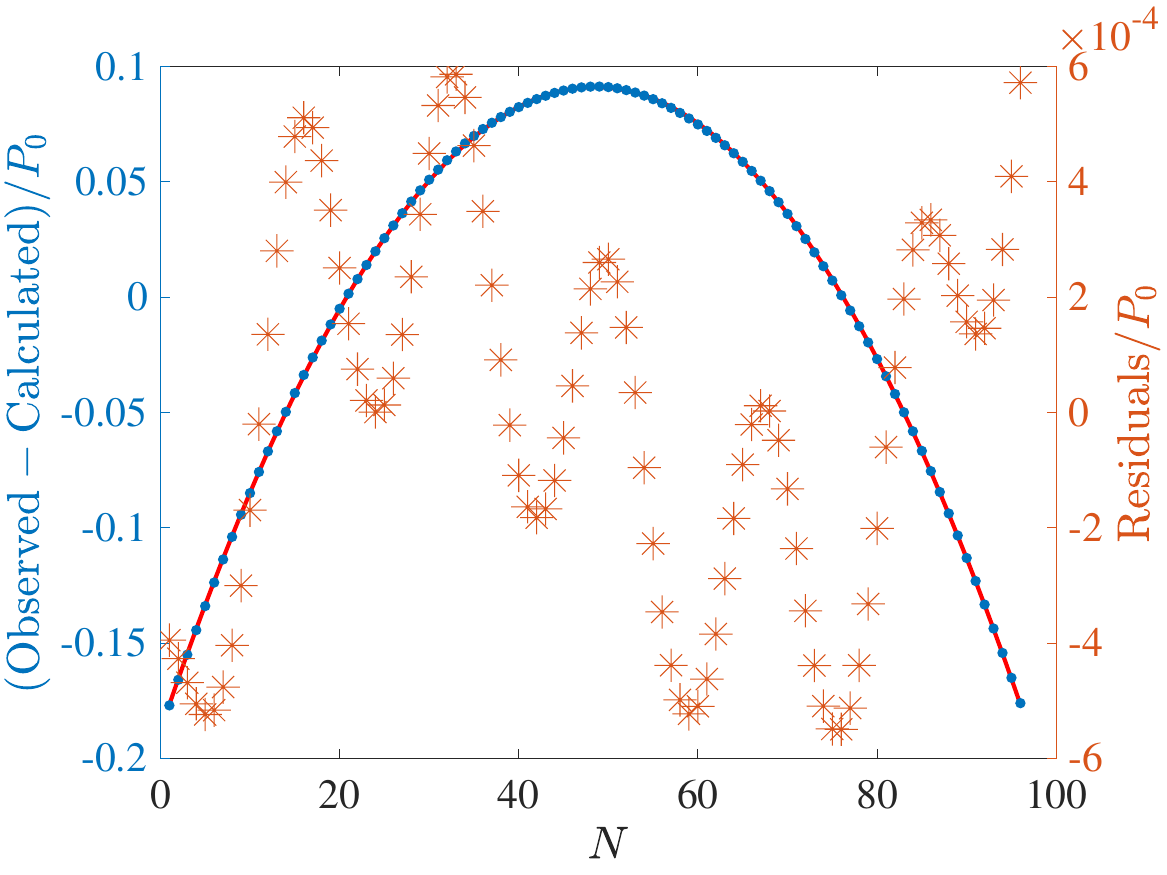}
     \end{subfigure}
     \hfill
     \begin{subfigure}[b]{0.5\textwidth}
         \centering
         \includegraphics[width=\linewidth]{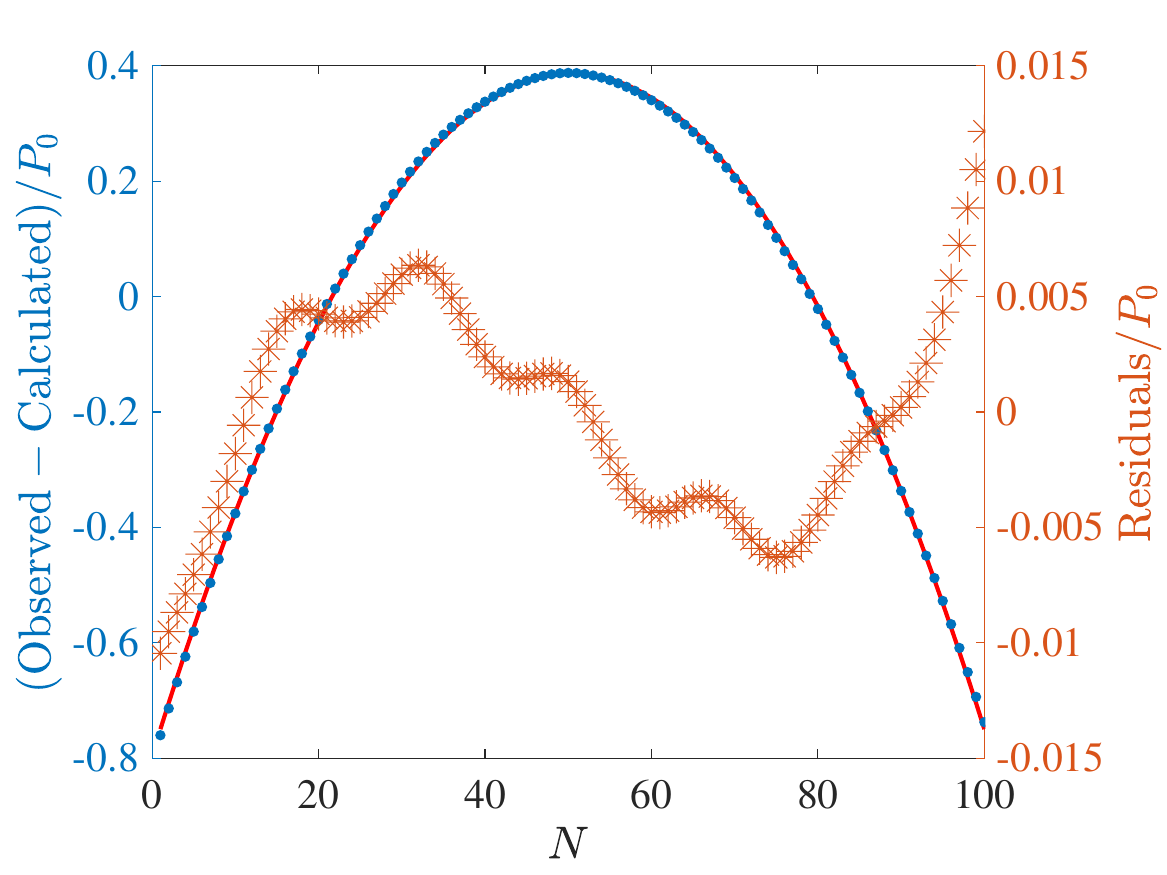}
     \end{subfigure}
        \caption{An example demonstrating the period evolution of a star interacting with an accretion disc. A disc with $\dot{m}/\dot{m}_{\rm Edd} = 10^{-2}$ and $\alpha = 10^{-2}$ and a star on an eccentric orbit $e=0.95$ orbiting an SMBH with $R_g/a = 10^{-3}$. The fitted linear ephemeris is subtracted from the recorded pericenter passage times, and plotted as blue dots. The residuals from a quadratic fit (solid red curve) are plotted as red asterisks (right hand side vertical axis). The residual has a distinct cubic component corresponding to $\ddot{P}$ as well as an additional oscillatory component reflecting the orbit's apsidal precession, causing the star to periodically interact with regions of varying surface density within the accretion disc. Top panel - $m_1=1$ and $R_1=1$, bottom panel - $R_1 = 2$.}
    \label{fig:Numerical_Pdot}
\end{figure}

To illustrate the effect, we integrate the star's orbital motion in the presence of a Paczyński–Wiita potential to capture the orbit's GR apsidal precession, and include the impulsive effect of drag (equation \ref{eq:deltaV_drag}), introduced whenever the star crosses the reference disc plane, in both ascending and descending nodes. A standard, steady accretion disc of $\dot{m}/\dot{m}_{\rm Edd} = 10^{-2}$ and $\alpha = 10^{-2}$ is assumed. The times of pericenter passages are recorded, and are fitted to a linear ephemeris, which is then subtracted. The O-C data are plotted in figure \ref{fig:Numerical_Pdot}. As expected, an average orbital period decay results in a distinct quadratic trend. Also plotted are the residuals from a quadratic fit to the O-C data, representing deviations from constant $\dot{P}$. The residuals are composed of a cubic trend related to $\ddot{P}$ (naturally arising from the next order perturbative term in the orbital evolution), as well as an oscillatory component arising from the orbit's apsidal precession, which gradually changes the radial location of the star-disc collisions. Motivated by the inferred orbit of ASASSN-14ko, we take $R_g/a \approx 10^{-3}$ and an initial orbit with $e=0.95$ in Figure \ref{fig:Numerical_Pdot}. The calculation is repeated for a star with $R_1=1$ and $R_1=2$ (top and bottom panels of figure \ref{fig:Numerical_Pdot}). The inferred period evolution is $\dot{P} \approx -2.4\times10^{-4}$ for $R_1=1$ and $\dot{P}=-9.6\times10^{-4}$ for $R_1=2$ (with a ratio of 4 between the two, as expected from equation \ref{eq:Pdot_avg_rp}).
The period of the oscillatory component is $T_{\rm prec,ap}/2$ (equation \ref{eq:T_prec_ap}), equivalent to $\sim 16$ orbits. {Note that the oscillatory modulations in $\dot{P}$ are not captured in \cite{Macleod_Lin_2020} as they consider the limit of Newtonian gravity, and thus inhibit precession.}

\section{Interpretation of the period evolution of ASASSN-14ko as disc-induced drag} \label{sec:14ko_Pdot_application}

Hydrodynamical drag acting during star-disc collisions results in an orbital period decay at an average rate given by equation \ref{eq:Pdot_avg_rp}. Given the observed value of $\dot{P}=2.6\times 10^{-3}$ of ASASSN-14ko, the constrained quantity is thus
\begin{multline} \label{eq:Sigmad_over_r_obs}
    \frac{\Sigmad(\rp)}{\rp} \approx \frac{m_\star}{3\pi R_{\star}^2 a} \dot{P} \approx \\ 7.6\times 10^{-8} \; m_1 R_1^{-2} \pfrac{\dot{P}}{2.6\times 10^{-3}} M_{\bullet,7}^{-1/3} \mathcal{P}_{115}^{-2/3} \; \rm g \, cm^{-3} \,,
\end{multline}
or
\begin{multline} \label{eq:Sigmad_from_Pdot1}
    \Sigmad = \frac{m_\star}{3\pi R_\star^2} \dot{P} \pfrac{\rp}{a} \approx \\
    1.1 \times 10^8 \; \pfrac{\rp}{a} R_1^{-2} m_1 \pfrac{\dot{P}}{2.6\times 10^{-3}} \; \rm g \, cm^{-2} \,,
\end{multline}
and considering $\rp \gtrsim \rpmin \approx 2 \times \rtidal$, the observed $\dot{P}$ sets a lower limit on the disc's surface density
\begin{multline} \label{eq:Sigmad_from_Pdot}
    \Sigmad \geq \frac{1}{3\pi} \frac{\MBH^{1/3} m_\star^{2/3} }{R_\star a} \dot{P} \pfrac{\rpmin}{2 \rtidal} \approx \\
    2.3 \times 10^6 \; R_1^{-1} m_1^{2/3} \mathcal{P}_{115}^{-2/3} \pfrac{\dot{P}}{2.6\times 10^{-3}} \pfrac{\rpmin}{2 \rtidal} \; \rm g \, cm^{-2} \,,
\end{multline}
corresponding to a minimal disc mass around $r\approx \rp$ of order
\begin{multline} \label{eq:mdisc_min}
    m_{\rm d} \approx \pi \rp^2 \Sigmad = m_\star \frac{\dot{P}}{3} \pfrac{a}{R_\star}^2 \pfrac{\rp}{a}^3 \geq \frac{8}{3} \MBH \dot{P} \frac{R_\star}{a} \pfrac{\rpmin}{2 \rtidal}^3 = \\
    = 3.2 \; M_{\bullet,7}^{2/3} R_1 \mathcal{P}_{115}^{-2/3} \pfrac{\dot{P}}{2.6\times 10^{-3}} \pfrac{\rpmin}{2 \rtidal}^3 \; \rm M_\odot \,.
\end{multline}
Remarkably, the above result does not directly depend on $m_\star$ nor on the disc scale height $h$. Yet, for main-sequence stars, $R_\star \propto m_\star^{0.8}$ and hence the ratio between the disc and stellar mass is nearly constant, $m_{\rm d}/m_\star \gtrsim 3$. For stars with extended radii, such that $R_1 \gg m_1$ (e.g., red giants), the disc mass certainly exceeds the mass of the star, $m_{\rm d}/m_\star \gg 1$.  
The possibility of a compact object and a cross-section set by gravitational focusing is presented in \ref{appendix:graviational_focusing}. However, given the observed $\dot{P}$, this regime would imply a disc mass greater by orders of magnitude relative to the case of a geometrical cross section. 

\subsection{Implication regarding the partial TDE interpretation}
\label{sec:implication}
ASASSN-14ko has been previously interpreted as a repeating TDE, grazing the tidal radius at pericenter passage, with the star's outer layers undergoing partial stripping of mass, repeatedly. Under this interpretation, it seems natural to associate the extracted material as the source of the accretion disc with which the star interacts, acting to reduce its orbital period through hydrodynamical drag. However, given the observed $\dot{P}$, the minimal disc mass is likely to exceed $m_\star$ (e.g., equation \ref{eq:mdisc_min}). This would only be plausible if the star has been previously stripped of a considerable fraction of its mass that produced the required disc. It also requires that the viscous time in the disc is long (so that the stripped mass primarily builds up over time, rather than accretes), making it unclear if the observed flares can be powered by accretion of stripped stellar material.

Alternatively, the star could be interacting with a preexisting, long-lasting disc, whose origin is unrelated to stripping of mass from the star, for example, an AGN accretion disc, which is indeed observationally evident in ESO 253-G003. In this case, $\rp$ is not necessarily very close to $\rtidal$, and a somewhat less eccentric orbit is permitted. This would, however, necessitate a higher density disc and correspondingly larger disc mass (note the scaling with $\rp$ in equations \ref{eq:Sigmad_from_Pdot} and \ref{eq:mdisc_min}).

We previously assumed $\iota \sim \mathcal{O}(1)$ in deriving the minimal disc mass. Note that even if the disc forms from tidally stripped stellar material which is initially confined to the same plane as the star, the difference between the nodal (Lense-Thirring) precession rates of the stellar orbit (of semi-major axis $a$) and the disc material (of size $\sim \rtidal \ll a$) rapidly brings them out of mutual alignment (as long as they are both misaligned with the SMBH spin). Thus, the assumption of order-unity inclination is appropriate for both the partial-TDE and pre-existing disc pictures.

\subsection{Comparison with standard disc models}

Next, we explore the conditions for an accretion disc to match the observed $\dot{P}$ through its surface density. The surface density at radius $r$ of an $\alpha$-disc, steadily accreting at a rate $\dot{m}$ is given by
\begin{equation} \label{eq:Sigma_d_gen}
    \Sigma_{\rm disc} = \frac{\dot{m}}{3 \pi \Omega_k h^2 \alpha} \approx 
    \frac{1}{6\pi^2} \frac{\dot{m} P_0}{\alpha h^2} \pfrac{r}{a}^{3/2} \,,
\end{equation}
where $\alpha$ is the Shakura-Sunyaev $\alpha$-parameter. If the disc midplane is radiation pressure dominated 
\begin{equation} \label{eq:h_rad_disc}
    \left. h \right|_{\rm rad} \approx \frac{3}{2\eta} R_g \pfrac{\dot{m}}{\dot{m}_{\rm Edd}} \approx 2.2\times 10^{13} \; M_{\bullet,7} \pfrac{\dot{m}}{\dot{m}_{\rm Edd}} \, {\rm cm} \, ,
\end{equation}
where $\dot{m}_{\rm Edd} = 1.7\times 10^{25} \, \rm M_{\bullet,7} \; \rm g \, s^{-1}$ is the Eddington accretion rate assuming a radiative efficiency of $\eta = 0.1$. Plugging in to equation \ref{eq:Sigma_d_gen}
\begin{multline} \label{eq:Sigma_rad_disc}
    \Sigma_{\rm disc,rad} \approx \frac{2 \eta^2}{27\pi^2} \frac{\dot{m}_{\rm Edd} P_0}{\alpha R_g^2} \pfrac{\dot{m}}{\dot{m}_{\rm Edd}}^{-1} \pfrac{r}{a}^{3/2} \approx \\
    \approx 6\times 10^5 \; 
    \mathcal{P}_{115} \alpha_{-2}^{-1} M_{\bullet,7}^{-1} \pfrac{\dot{m}}{\dot{m}_{\rm Edd}}^{-1} \pfrac{r}{a}^{3/2} 
    \; \rm g \, cm^{-2} \,,
\end{multline}
where $\alpha_{-2} = \alpha/10^{-2}$. Alternatively, if the disc is gas pressure dominated at this radius
\begin{equation} \label{eq:h_gas_disc}
    \left. h \right|_{\rm gas} \approx \pfrac{3 k_{\rm B}^4}{64 \pi^2 \sigma_{\rm SB} \mu_{\rm p}^4}^{1/10} \pfrac{\kappa}{\alpha}^{1/10} \dot{m}^{1/5} r^{21/20} (G\MBH)^{-7/20} \,,
\end{equation}
where $\mu_{\rm p}$ is the proton mass, $\kappa \approx 0.34 \, \rm cm^2 \, gr^{-1}$ is the electron scattering opacity, $k_{\rm B}$ the Boltzmann constant and $\sigma_{\rm SB}$ is the Stefan-Boltzmann constant. The corresponding surface density is
\begin{equation} \label{eq:Sigma_gas_disc}
    \Sigma_{\rm disc, gas} \approx 4.4\times 10^6 \; \pfrac{\dot{m}}{\dot{m}_{\rm Edd}}^{3/5} \alpha_{-2}^{-4/5} \mathcal{P}_{115}^{-2/5} M_{\bullet,7}^{3/5} \pfrac{r}{a}^{-3/5} \; \rm g \, cm^{-2} \,.
\end{equation}

The maximal surface density (at fixed $r$) is achieved when gas and radiation pressure are comparable, yielding
\begin{multline} \label{eq:Sigmad_max}
    \Sigma_{\rm disc,max} = \frac{32}{15 \sqrt[8]{3} \sqrt[4]{5} } \left( a_{\rm rad}^{-6}
    \pfrac{\kb}{\mu_{\rm p}}^{-4} \pfrac{\sigma_{\rm SB}}{\kappa \alpha}^{7} \Omega_k^{-1} \right)^{1/8} \approx \\
    9\times 10^5 \; \alpha_{-2}^{-7/8} \mathcal{P}_{115}^{1/8} \pfrac{r}{a}^{3/16} \; \rm g \, cm^{-2} \geq \\
    4.3 \times 10^{5} \; \alpha_{-2}^{-7/8} \mathcal{P}_{115}^{3/4} R_1^{3/16} m_{1}^{-1/16} \pfrac{r_{\rm p,min}}{2\rtidal}^{3/16} \,,
\end{multline}
where $a_{\rm rad}$ is the radiation constant. 

At higher accretion rates the disc is dominated by radiation pressure (and is described by equations \ref{eq:h_rad_disc} and \ref{eq:Sigma_rad_disc}), while at lower accretion rates, gas pressure dominates (equations \ref{eq:h_gas_disc} and \ref{eq:Sigma_gas_disc}). {This estimate of the maximum disc surface density is uncertain at the order of magnitude level because it neglects magnetic support, vertical transport of energy by turbulence, and relies on radiation dominated disc solutions that are known to be thermally and viscously unstable \citep{Lightman74,Shakura1976}.  Indeed, the existence of a maximum disc surface density is precisely because $\Sigmad \propto \dot m^{-1}$ in the radiation dominated regime, which is the origin of the viscous instability of radiation dominated discs.}

Note that the disc surface density as implied by the observed $\dot{P}$ scales as $\Sigmad \propto \rp$ (equation \ref{eq:mdisc_min}), whereas the maximal $\alpha$-disc scales as $\rp^{3/16}$, suggesting that the required conditions may only prevail at sufficiently small $\rp$. As a consequence, we conclude that $\rp$ cannot be too large, or else the disc would not be massive enough to decrease the orbital period at the observed rate.

Given the quiescent X-ray emission of $L_Q \approx 10^{43} \, \rm erg \, s^{-1}$, the disc's Eddington ratio is constrained to within an order of magnitude $\dot{m}/\dot{m}_{\rm Edd} \approx 8\times 10^{-3} M_{\bullet,7}^{-1}$. The limit on the quiescent luminosity and the disc's accretion rate
favors a gas-pressure dominated disc, occurring at lower Eddington ratios. 

\begin{figure}
    \centering
    \includegraphics[width=0.5\textwidth]{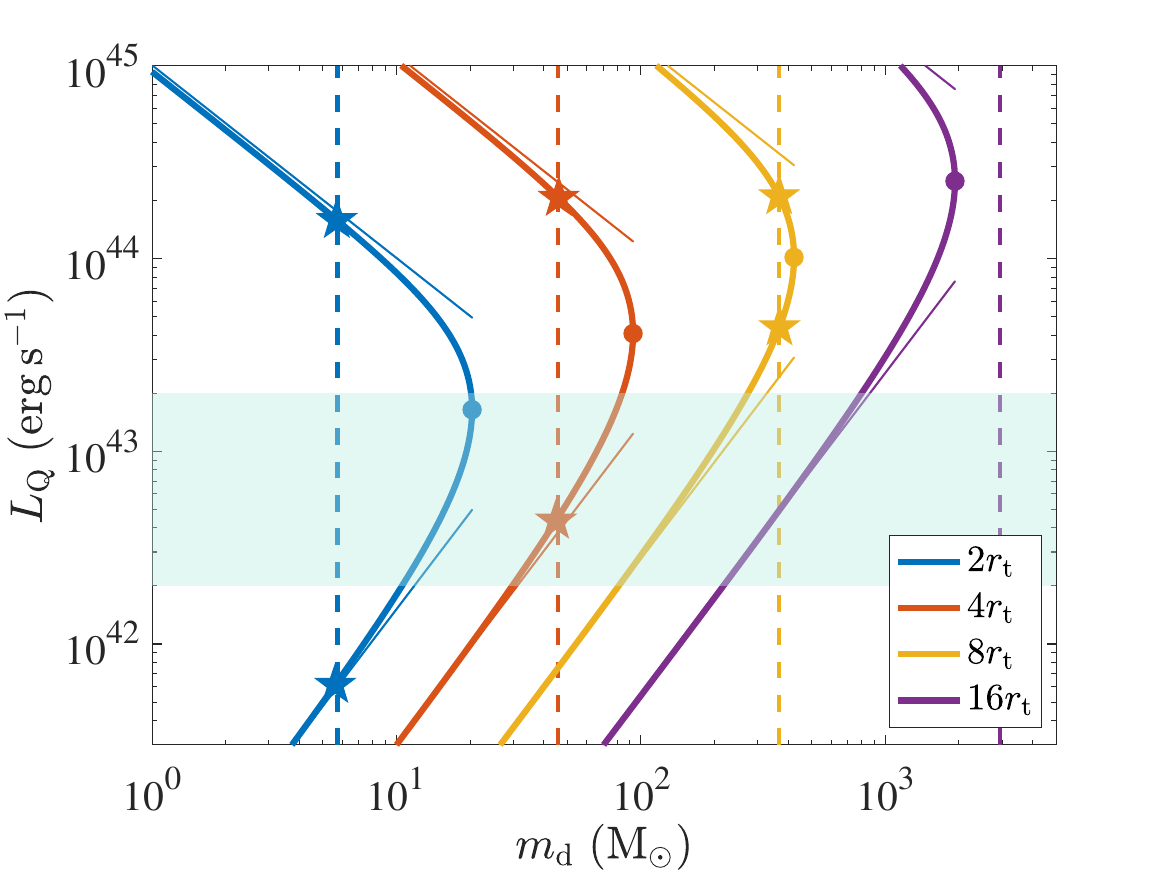}
    \caption{Accretion disc properties and constraints from ASASSN-14ko's orbital evolution. Solid lines correspond to standard steady-accretion disc models around an SMBH of mass $\MBH = 10^7 \, \rm M_\odot$ with $\alpha = 10^{-2}$ at a few different radii. Plotted is the disc's bolometric luminosity $L_{\rm Q}$ (closely related to its accretion rate, $L_{\rm Q}\approx \eta \dot{m} c^2$, $\eta\approx 0.1$) as a function of the enclosed disc mass at a given radius $r$ (closely related to the disc's surface density, $\Sigmad$). The disc solution has two branches corresponding to a gas-pressure or a radiation-pressure dominated disc, delineated by a maximal disc mass represented by a solid circle. The two thin lines show the asymptotic behavior given in equations \ref{eq:Sigma_rad_disc} and \ref{eq:Sigma_gas_disc}. The vertical dashed lines show the disc mass inferred from the orbital period decay measured for ASASSN-14ko, $\dot{P}\approx 2.6\times 10^{-3}$, assuming a star of mass $m_\star = \rm M_\odot$ and radius $R_\star = {5} \, \rm R_\odot$. Solid stars show the disc solutions which produce the observed $\dot{P}$. The shown radii correspond to the star's pericenter distances, taken at multiples of the star's tidal radius, $\rtidal \approx {5} \; \rm au$. The shaded region shows the bolometric luminosity of the accretion disc, as constrained by ASASSN-14ko's X-ray observations.}
    \label{fig:Luminosity_vs_disc_mass}
\end{figure}

These arguments are summarized in Figure \ref{fig:Luminosity_vs_disc_mass}, showcasing the quiescent disc luminosity as a function of its mass at different radii, $\rp/\rtidal = 2,4,8,16$. The solid curves represent steady-state accretion discs with $\alpha=10^{-2}$, over a wide range of accretion rates, where the disc midplane pressure is dominated by either gas (at low $L_{\rm Q}$, equation \ref{eq:Sigma_gas_disc}) or radiation (at high $L_{\rm Q}$, equation \ref{eq:Sigma_rad_disc}). Each curve is characterized by a maximal disc mass, given by $m_{\rm d, max} = \pi \rp^2 \Sigma_{\rm disc, max}$ (solid circle). For each radius $\rp$, the disc mass necessary to explain the observed $\dot{P}$ (i.e., equation \ref{eq:mdisc_min}) is plotted as vertical dashed line, assuming a star with $m_1 = 1$ and $R_1 = 5$. The disc's quiescent luminosity (of roughly $10^{42}-10^{43} \, \rm erg \, s^{-1}$) is shown as a shaded horizontal strip.

\begin{figure}
    \centering
    \includegraphics[width=0.5\textwidth]{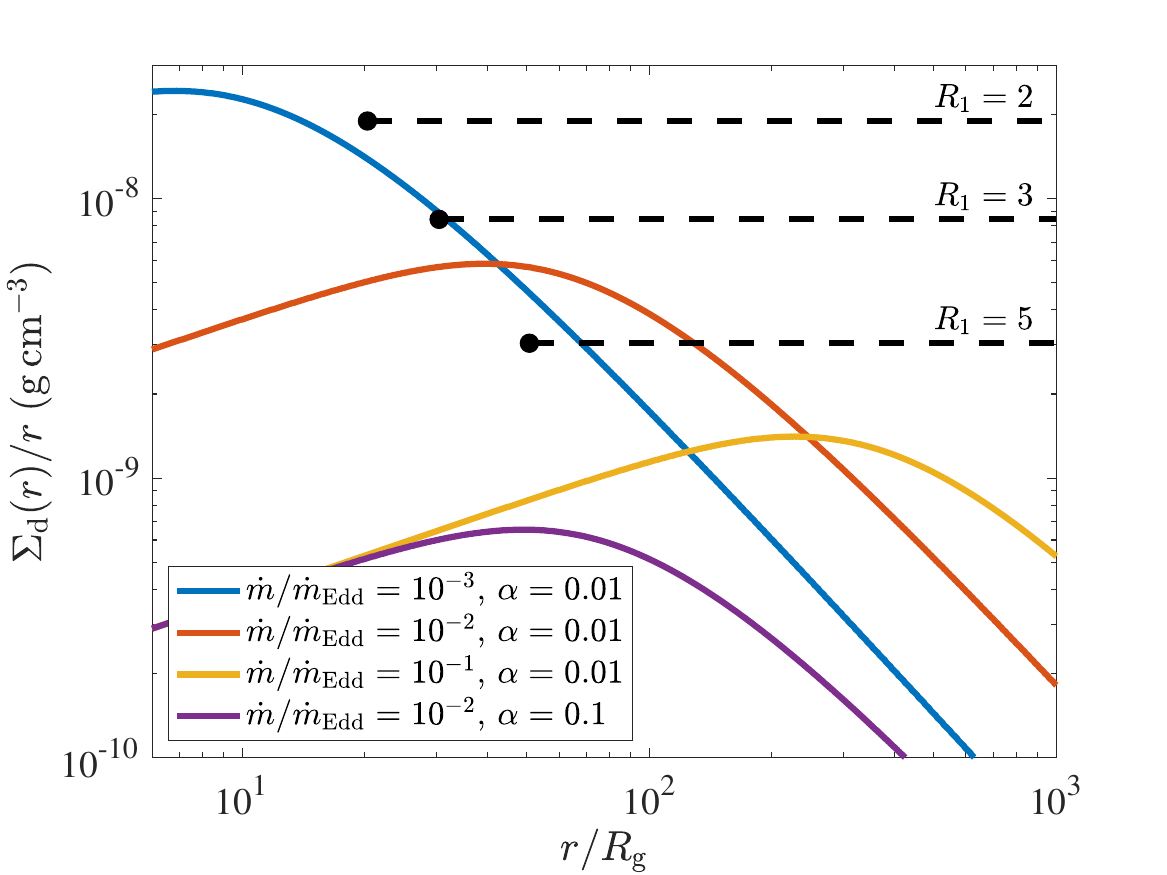}
    \caption{Standard accretion disc profiles and constraints implied by ASASSN-14ko. Colored plots correspond to different disc parameters, around an SMBH of $\MBH=10^7 \, \rm M_\odot$, plotting $\Sigmad(r)/r$ as a function of radius. The horizontal dashed lines correspond to the value of $\Sigmad/r$ at pericenter required to generate the observed $\dot{P}$ of ASASSN-14ko, for different values of stellar radii, $R_1=R_\star/\rm R_\odot$ and mass $m_\star = 1 \, \rm M_\odot$. Horizontal lines are terminated at the tidal radius (solid circle).}
    \label{fig:ThinDisc}
\end{figure}

Another demonstration of the relation between the disc properties and the observed period evolution is shown in figure \ref{fig:ThinDisc}. A number of standard thin-disc models are plotted, for representative values of $\dot{m}$ and $\alpha$. For different stellar radii, $R_1 = 2,3,5$, the value of $\Sigmad(\rp)/\rp$ consistent with ASASSN-14ko (equation \ref{eq:Sigmad_over_r_obs}) is plotted as a dashed horizontal line, where a solution for the observed orbital decay exists where this line intersects with a solid colored curve profile. The dashed horizontal lines are terminated at $\rtidal$. A solution is only marginally achieved for some of the disc models, and generally requires that the star is somewhat inflated, with $R_1 \approx$ few. {Figure \ref{fig:ThinDisc} also demonstrates that larger values of $\alpha$ ($\gg 10^{-2})$ are disfavored, as they correspond to an overall lower $\Sigmad$, which cannot produce the observed $\dot{P}$ for reasonable stellar mass and radius (compare purple and red curves).}

The observational constraints ($L_{\rm Q,obs}$ and $\dot{P}$) imply a fairly limited permitted parameter space: (a) the disc is likely gas pressure dominated, accreting at roughly $\dot{m} \approx 10^{23} \, \rm g \, s^{-1}$; (b) the star is somewhat inflated relative to a main-sequence star, likely due to tidal heating\footnote{{The alternative, that the star is an evolved subgiant, is not tenable.  The reason is that the pressure of the inflated outer radii in a subgiant is orders of magnitude smaller than the ram pressure due to interaction with the disc (see section \S \ref{sec:ablation} and equations \ref{eq:pram} and \ref{eq:pstar}); these layers will get stripped but are very unlikely to contribute to the effective area that matters for drag.  By contrast, stars inflated by tidal heating in their interior have larger pressures at large radii relative to a subgiant of the same radius.}}; (c) its pericenter distance is just a few times greater than its tidal radius, $\rp \gtrsim 2\rtidal \approx 10 \, \rm au$ (and $e \lesssim 0.9$); (d) the cumulative disc mass at this radius is of order 10 solar masses.  

\subsection{Stellar mass loss and the origin of ASASSN-14ko's flares} \label{sec:ablation}
In addition to dissipating the orbital period, star-disc interaction may also affect the stellar structure and ablation. If the observed $\dot{P}$ is indeed the result of collisions with a disc, the ram pressure experienced by the star is approximately
\begin{multline} \label{eq:pram}
    p_{\rm ram} \approx \frac{1}{2} \rhod v_k^2 \approx \frac{\Sigmad}{4h} \frac{2G\MBH}{\rp} = \frac{2}{3} \frac{G m_\star^2}{4\pi R_\star^4} \frac{\MBH R_\star^2}{m_\star a h} \dot{P} = \\
    5.6\times 10^{14} \; \pfrac{\dot{P}}{2\times10^{-3}} \pfrac{h}{\rm R_\odot}^{-1} R_1^{-2} m_1 \mathcal{P}_{115}^{-2/3} M_{\bullet,7}^{2/3} \; \rm erg \, cm^{-3} \,,
\end{multline}
where we subsituted $\Sigmad/\rp$ from equation \ref{eq:Sigmad_over_r_obs}. For our fiducial values, the ram pressure is comparable to the star's interior mean pressure 
\begin{equation}
    p_\star \approx Gm_\star^2 / (4\pi R_\star^4) \approx 9\times 10^{14} \; R_1^{-4} m_1^2 \; \rm erg \, cm^{-3} \,.
    \label{eq:pstar}
\end{equation}
By interpolating the results found in the high-resolution simulations of \cite{Liu_2015} to arbitrary stellar mass and radius, we find that the mass ablated from the star per disc passage is
\begin{multline}
    \delta m \approx 5\times 10^{-3} \; m_\star \pfrac{p_{\rm ram}}{p_\star} \approx \\
    3\times 10^{-3} \; {\rm M_\odot} \; \pfrac{\dot{P}}{2\times10^{-3}} \pfrac{h}{\rm R_\odot}^{-1} R_1^{2} \mathcal{P}_{115}^{-2/3} M_{\bullet,7}^{2/3} \,.
    \label{eq:dM_ablation}
\end{multline}
{We note, however, that equation \ref{eq:dM_ablation} does not apply in the limit where the shock duration is less than the dynamical time of the star, as is the case in our problem because $h/r \ll 1$ for the disc.  This will suppress the mass ablated relative to that predicted by equation \ref{eq:dM_ablation}}.  

{Equation \ref{eq:dM_ablation} poses yet another constraint on the radius of the orbiting star. The survival of the star over many ($\gg 20$) orbit, implies that $R_1 \lesssim 4$, or else the mass loss through ablation becomes too large. A significantly inflated star (such as a giant) faces another related challenge - once $p_{\rm ram} \gg p_\star$, the effective cross section of the star contributing to the drag is diminished to less than $R_\star$, and the resulting $\dot{P}$ decreases accordingly.}

Given the uncertainties, the mass-loss per orbit predicted by equation \ref{eq:dM_ablation} is of order what is required to explain the flares in ASASSN-14ko.  In particular, the observed radiated flare energy could be explained by the accretion of roughly $10^{-3} \, \eta_{0.1}^{-1} \, \rm M_\odot$, as discussed in \S \ref{sec:general_emission}.

If, on the other hand, mass loss from the star occurs due to tidal stripping of the star's outer layers, inflated by shock heating, as proposed in \cite{Lu_Quataert_2023}, mass lost from the star per pericenter passage is roughly
\begin{equation}
    \delta m \approx m_\star \left( \frac{\dot{P}}{3} \frac{\MBH R_\star^2}{m_\star a h} \right)^{K} \,,
\end{equation}
{where $K$ depends on the stellar envelope's density profile - $K=7/5$ for a convective envelope and $K=5/4$ for a radiative envelope \citep[see appendix B of][]{Lu_Quataert_2023}. For fiducial values and assuming $K=7/5$
\begin{equation}
    \delta m \approx 0.2 \, {\rm M_\odot} \; \pfrac{h}{\rm R_\odot}^{-7/5} \pfrac{\dot{P}}{2\times10^{-3}}^{7/5} M_{\bullet,7}^{14/15} R_1^{14/5} \mathcal{P}_{115}^{-14/15} m_1^{-2/5} \,,
    \label{eq:dMRLO}
\end{equation}
much greater than the mass lost due to ablation, and uncomfortably large given that at least $\sim$20 flares have occurred. We note, however, that equation \ref{eq:dMRLO} has not been calibrated using simulations so there is a significant uncertainty in the correct normalization of this expression.   

The disc mass directly impacted by the star is given by
\begin{multline}
    m_{\rm sh} \approx \Sigmad \pi R_\star^2 = \frac{1}{6} m_\star \dot{P} \pfrac{\rp}{a} = \\
    3.3\times 10^{-4} \; \pfrac{\rp}{a} m_1 \pfrac{\dot{P}}{2\times 10^{-3}} \; \rm M_\odot \,,
\end{multline}
and given $\rp \gtrsim r_{\rm p,min}$, the minimal disc mass shocked by the star is
\begin{equation}
    m_{\rm sh} \gtrsim 6.7\times10^{-6} \; R_1 m_1^{2/3} \pfrac{r_{\rm p,min}}{2 \rtidal} \pfrac{\dot{P}}{2\times 10^{-3}} \mathcal{P}_{115}^{-2/3} \; \rm M_\odot \,.
\end{equation}
The disc mass directly shocked by the star each orbit is significantly less than the mass stripped from the star by the star-disc interactions.  In the context of our models, this suggests that the star's envelope is the reservoir of fuel powering the observed flares in ASASSN-14ko, rather than the impacted disc material. 

{If indeed ablation due to star-disc collisions ultimately powers the flares, the two disc passages per orbit may result in \textit{two} rather than a single flare per orbit, if the disc extends to radii $\gg \rp$. These two flares might overlap and appear as one if their duration $\Delta t$ exceeds the time between the two disc crossings occurring before and after pericenter passage. If the flare duration is longer than the pericenter passage time, $\Delta t \gg t_{\rm p} \approx P_0 (1-e)^{3/2}$, this condition is only satisfied when the orbit's major axis is roughly within an angle $\theta_c \approx (t_{\rm p}/\Delta t)^{1/3}$ of the disc plane. As the orbit precesses, two distinct flares per orbit temporarily appear over the course of $\Delta N \lesssim T_{\rm prec,ap}/P_0$ orbits, with}
\begin{multline}
    \Delta N \approx \pfrac{t_{\rm p}}{\Delta t}^{1/3} \frac{\rp}{R_g} \approx \\
    3 \; \pfrac{\rp}{2 \rtidal}^{3/2} R_1^{3/2} m_1^{-1/2} \pfrac{\Delta t}{10 \, \rm d}^{-1/3} M_{\bullet,7}^{-2/3} \,,
    \label{eq:2flares}
\end{multline}
before merging back to a single distinct flare for the remaining precession cycle (lasting $T_{\rm prec,ap}/P_0 \approx 13 \, (\rp/2\rtidal)$ orbits). {In this regime, the two disc passages occur at considerably different radii, $\rp \leq r_1 \ll r_2 \leq r_{\rm max}$, with the outer disc passage resulting in weaker shocks, smaller amount of ablated mass, and consequently, weaker flare relative to the inner collision site. Thus, even when the two disc crossings are well-separated in time, only one bright flare per orbit may be observed.}

\section{Other sources of period evolution - ASASSN-14ko}
\label{sec:other}
In this section we consider a few other potential origins of  observed period evolution in periodic nuclear transients (PNTs) - gravitational wave emission, light travel time and orbital precession or barycentric acceleration, tides, and mass transfer.  We show that all are ruled out for ASASSN-14ko, being inconsistent with the observations and/or requiring unreasonable physical conditions to realize. 

\citet{Payne_2021} suggested that the orbital $\dot P$ of ASASSN-14ko is broadly consistent with the energy change imparted to the surviving stellar core in simulations of partial stellar disruptions (e.g., \citealt{Ryu2020}).  The orbital decay found in \citet{Ryu2020}'s partial disruption simulations is due to tidal energy transfer from the orbit to the star (see also \citealt{Cufari2023}).   We show in \S \ref{sec:tides} and \S \ref{sec:masstfer} that this explanation for the orbital decay in ASASSN-14ko faces severe challenges.

\subsection{Gravitational Waves}
Gravitational wave emission leads to orbital period decays at a rate \citep[e.g.,][]{Peters_1964}
\begin{multline}
    \left| \dot{P}_{\rm GW} \right| \\
    \approx \pfrac{85 \pi \sqrt{2}}{8} \frac{G^{5/2} \MBH^{3/2} m_\star a}{c^5 \rp^{7/2}} \leq
    10^{-6} \; \mathcal{P}_{115}^{3} M_{\bullet,7}^{11/6} m_1^{13/6} R_1^{-7/2} \,,
\end{multline}
where the above limit corresponds to $\rp\geq \rtidal$. This result is several orders of magnitude lower than the observed value \citep{Payne_2021,Cufari_2022}. Even if a more compact stellar object is considered (taking $R_\star \lesssim \rm R_\odot$), the observed $\dot{P}$ still cannot be fulfilled by gravitational waves, since $\rp$ must be at least a few times greater than $R_g$, limiting $\dot{P}_{\rm GW} \lesssim 10^{-5}$. 

\subsection{Flare timing traces star-disc collisions - apsidal precession} \label{sec:Precession_modulation}

If the flares' timing is directly determined by the star-disc collision schedule (rather than say, pericenter passages), relativistic apsidal precession may result in an apparent $\dot{P}$, due to a mostly geometric effect. Assuming that the plane of the disc remains fixed, the time between two consecutive collisions is determined by the apsidal precession angle acquired over one orbital period,
\begin{equation}
    \Delta \varphi = \frac{2\pi P_0}{T_{\rm prec,ap}} \approx \frac{6\pi}{1-e^2} \frac{R_g}{a} \,,
\end{equation}
such that
\begin{equation}
    \delta t (\theta) = t_{i+1}-t_i \approx P_{\rm orb} \left( 1 \pm \frac{r^2(\theta) \Delta \varphi }{2 \pi a^2 (1-e^2)} \right) \,,
\end{equation}
where collisions occur at true anomaly $\theta$ measured with respect to the orbit's periapsis, and at which the radial distance from the SMBH is $r(\theta) = a(1-e^2) / (1+e \cos{\theta})$. The sign of the second term is determined by the orbit's sense of rotation. The corresponding time derivative is then given by
\begin{equation}
    \dot{P} \approx \frac{\partial (\delta t (\theta) )}{\partial \theta} \frac{\Delta \varphi}{P} \,,
\end{equation}
and after some algebra
\begin{equation}
    \dot{P} \approx 36\pi \pfrac{R_g}{a}^2 \frac{e \sin{\theta}}{(1-e^2) (1+e \cos{\theta})^3} \,,
\end{equation}
For fiducial values
\begin{equation} \label{eq:Pdot_from_prec_sp}
    \dot{P} \approx 10^{-4} \frac{e \sin{\theta}}{(1-e^2) (1+e \cos{\theta})^3} \; \mathcal{P}_{115}^{-4/3} \; M_{\bullet,7}^{4/3}  \,.
\end{equation}

Note that the period derivative changes sign along the orbital precession timescale. The second time derivative is given by
\begin{equation} \label{eq:Pddot_geo}
    \ddot{P} \approx \frac{\partial \dot{P}}{\partial \theta} \frac{\Delta \varphi}{P} = \frac{216 \pi^2 e}{(1-e^2)^2} \pfrac{R_g}{a}^3 \frac{\cos{\theta} + e \cos^2{\theta} + 3 e \sin^2{\theta}}{(1+e \cos{\theta})^4} \frac{1}{P} \,,
\end{equation}
and for fiducial values
\begin{equation}
    \ddot{P} \approx 2\times 10^{-13} \, \frac{e(\cos{\theta} + e \cos^2{\theta} + 3 e \sin^2{\theta})}{(1-e^2)^2 (1+e \cos{\theta})^4} \; \mathcal{P}_{115}^{-3} M_{\bullet,7}^{4} \; \rm s^{-1}\,.
\end{equation}

ASASSN-14ko's $\dot{P}$ behavior can be explained by an orbital eccentricity $e\gtrsim 0.75$, when the collision angle is roughly $\theta \gtrsim \pi$, and the collision radius is roughly $r_2 \approx r_{\rm max}/2 \approx a$. At this radius, $|dr(\theta)/d\theta|$ is maximal, producing the desired effect. However, the lack of substantial evolution in $\dot{P}$ over a decade of observations implies that $|\ddot{P}| \ll \dot{P}_{\rm obs} / (10 \, \rm yr) \approx 6 \times 10^{-12} \; \rm s^{-1}$. These constraints are summarized in figure \ref{fig:Pdot_from_Precession}, demonstrating the incompatibility of this effect as the origin of ASASSN-14ko's timing behavior. Another challenge with this interpretation is that, with $e \approx 0.75$, the dynamical time near pericenter is quite long, of order a week.   It is not clear that one can produce flares with a duration also of order a week from such an orbit.  

\begin{figure}
    \centering
    \includegraphics[width=0.5\textwidth]{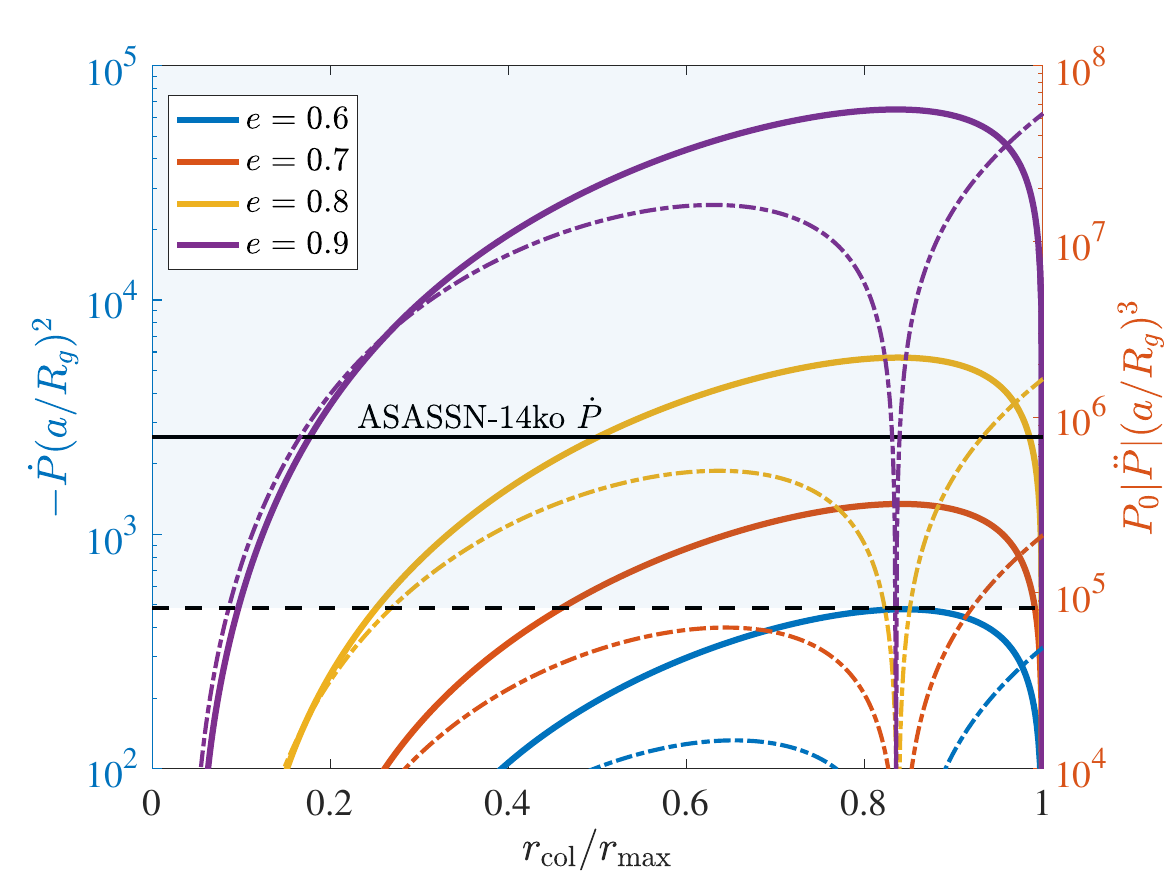}
    \caption{The period derivative resulting from relativistic apsidal precession and collisions with a disc midplane, as given by equation \ref{eq:Pdot_from_prec_sp}. The horizontal axis is the collision radius, relative to the apocenter distance $r_{\rm max}$. The left vertical axis is the (negative) period derivative of the timing of the flares, shown in solid curves.  The right vertical axis is proportional to $|\ddot{P}|$, normalized appropriately, plotted in dashed curves (equation \ref{eq:Pddot_geo}). Different colors correspond to different eccentricity values, as shown in the legend. The black horizontal line corresponds to the observed $\dot{P}$ of ASASSN-14ko, assuming $\MBH = 10^7 \, \rm M_\odot$. The shaded area above the dashed horizontal line is excluded by the lack of significant change in $\dot{P}$ over the course of ~10 years, thus ruling out disc collisions and apsidal precession as a source of the observed $\dot{P}$.}
    \label{fig:Pdot_from_Precession}
\end{figure}

\subsection{Light Travel Time Effects}

If the position of the region emitting the flares is systematically moving away or towards the observer on Earth, then the change in light travel-time from the flare-emitting region to the observer can manifest as a $\dot P$.  We consider two possibilities:   precession of the flare emitting region and barycentric acceleration (e.g., if the SMBH that the companion is orbiting is itself in a binary accelerating towards or away from Earth). 

\subsubsection{Precession}

If the center of light of the flare emitting region is precessing towards the Earth, light-travel time effects could manifest as an observed $\dot P < 0$.  Such precession of the light-emitting region could occur, e.g., if flares are produced by stream-stream intersections of tidal debris whose orbits precess due to apsidal or Lense-Thirring precession along with the orbiting companion.     Alternatively, precession of the companion's orbit could be produced by another object in the system via Kozai-Lidov interactions (as proposed in \citealt{King_2023} for ASASSN-14ko).    

A simple argument demonstrates that orbital precession of the center of light together with light travel-time effects cannot  explain the $\dot P$ observed in ASASSN-14ko.  Assume that the orbit precesses by an amount $\Delta \varphi$ per orbit and that the optical emission is dominated by gas at a radius $r_{e}$ ($\sim 10^{15}$ cm in ASASSN-14ko; \citealt{Payne_2021,Payne_2022}).   The change in physical distance between the observer and the emitting region satisfies $\Delta r \lesssim \Delta \varphi R_e$ and thus the implied change in period per orbit is given by $\Delta P \lesssim f \Delta \varphi R_e/c$.   The factor $f \le 1$ accounts for the emission geometry:  $f \simeq 0$ if the emission is spherically symmetric (so there is no change in the position of the center of light in spite of the precession) while $f \simeq 1$ is the emission is maximally asymmetric, e.g., dominated by gas on one side of the black hole.  The observed $\Delta P = \dot{P} P \simeq 7$ hrs of ASASSN-14ko, together with the observed blackbody emission radius $R_e \simeq 10^{15}$ cm thus imply $\Delta \varphi$ (per orbit) $\gtrsim 0.6/f$ rad per orbit.   For $f \sim 1$, this value is not implausible for apsidal precession.    However, precession and light-travel time effects as an explanation for the observed $\dot P$ thus generically predicts that the sign of $\dot P$ should switch due to precession by $\sim 2 \pi$ after $\sim 10 f$ orbits, which is inconsistent with the constant $\dot P$ observed by \citet{Payne_2022}, over $\sim 20$ orbits so far.   The only way out of this conclusion is to argue that the observed optical emission comes from a distance from the black hole much larger than the observed blackbody emission radius.     

The argument given here does not apply to the proposed origin of $\dot P$ via precession in \S \ref{sec:Precession_modulation}.   In that case precession by a given angle per orbit produces a significantly larger change in flare arrival time  (and hence a larger $\dot P$) than light travel time alone would produce.  This is because the star is orbiting at a speed that is much smaller than the speed of light. 

As a concrete example, we consider the relativistic apsidal precession angle associated with ASASSN-14ko
\begin{equation}
    \Delta \varphi = \frac{6\pi}{1+e} \frac{R_g}{\rp} \lesssim 0.9 \; R_1^{-1} m_1^{1/3} M_{\bullet,7}^{2/3} \,,
\end{equation}
where we have taken $\rp \gtrsim \rtidal$, hence $\Delta \varphi$ is marginally large enough to contribute substantially to the the observed $\dot{P}$ (if $R_\star$ is taken to be small enough). However, the sign of $\dot{P}$ should have switched twice every $(2\pi/\Delta \varphi)P \approx 8 \, \rm yr$, yet the observations point to a constant, negative $\dot{P}$ over the past 8 years. This implies that the optical emitting region must be relatively spherically symmetric so that the center of light position does not change significantly even if the stellar orbit is undergoing significant apsidal precession.
\end{comment}

\subsubsection{Barycentric acceleration}
An observed $\dot{P} < 0$ would also occur if the system as a whole is accelerating towards the Earth with a line-of-sight acceleration $a_{\rm acc}$. This would imply a barycentric acceleration of order
\begin{equation}
    a_{\rm acc} \approx \frac{\dot{P} c}{P} \approx 8 \; \mathcal{P}_{\rm 115}^{-1} \pfrac{\dot{P}}{2.6\times 10^{-3}} \; \rm cm \, s^{-2} \,,
\end{equation}
exceeding the Earth-sun acceleration by about an order of magnitude.

Although we do not know of a general argument that rules this out in ASASSN-14ko, the required acceleration is very large, much larger, e.g.,  than the acceleration provided by the second nucleus in the host galaxy ESO 253-G003.  
To see this, suppose that this acceleration is due to gravitational pull by an external mass $m_{\rm ext}$ at distance $r_{\rm ext}$. Since no significant variations in the observed $\dot{P}$ have been detected for $T_{\rm obs} = 8 \rm \, yr$, the orbital period must be longer than say $\sim 2 T_{\rm obs}$, and thus
\begin{equation}
    r_{\rm ext} > a_{\rm acc} (2 T_{\rm obs} )^2 / (4\pi^2) \approx 2600 \; \rm au \,,
\end{equation}
and the accelerating mass must exceed
\begin{equation}
    m_{\rm ext} \gtrsim \frac{a_{\rm acc} r_{\rm ext}^2}{G} \approx 10^8 \; \rm M_\odot \,.
\end{equation}
If we take $r_{\rm ext} \simeq 1.4$ kpc (the distance between the two nuclei in ESO 253-G003), the required mass is unphysically large, $m_{\rm ext} \simeq 10^{18} M_\odot$!  Thus the observed acceleration can only be explained if ASSASN-14ko is part of a binary black hole system with a separation of $\sim 3 \times 10^3-10^4$ au, contained within the NE nucleus of ESO 253-G003 (with the southern nucleus having yet another black hole given its observed AGN).  We regard this as unlikely.

\subsection{Tides}
\label{sec:tides}
Energy transfer between the orbit and the star through the action of tides is a natural source of observed orbital decay.  However, the orbital energy dissipation per orbit implied by the observations of ASASSN-14ko is given by
\begin{equation}
    \Delta E_{\rm orb} \approx \frac{2}{3} \frac{G \MBH m_\star}{2 a} \dot{P} \approx 10^{48} \; \pfrac{\dot{P}}{2.6\times10^{-3}} m_1 \mathcal{P}_{115}^{-2/3} M_{\bullet,7}^{2/3} \; \rm erg \,,
\end{equation}
which amounts to a large fraction of the star's binding energy
\begin{equation}
\label{eq:dEorb_obs}
    \frac{\Delta E_{\rm orb}}{E_\star} = 0.4 \; \pfrac{\dot{P}}{2.6 \times10^{-3}} M_{\bullet,7}^{2/3} \mathcal{P}_{115}^{-2/3} R_1 m_1^{-1} \,.
\end{equation}
where $E_\star = Gm_\star^2 / R_\star$.  It is unclear if tidal energies of this magnitude are even realizable; e.g., \citet{Cufari2023} find  a maximum tidal energy transfer per orbit of $\simeq 0.025 E_*$ for $n = 3/2$ polytropes in numerical simulations of partial TDEs.  Even if tidal energies this large are realizable, the star is unlikely to store this amount of tidal energy without being fully disrupted. Furthermore, the orbital decay rate corresponds to a luminosity
\begin{equation}
    | \dot{E}_{\rm orb} | = \frac{\Delta E_{\rm orb}}{P} \approx 10^{41} \; \pfrac{\dot{P}}{2.6\times10^{-3}} m_1 \mathcal{P}_{115}^{-5/3} M_{\bullet,7}^{2/3} \; \rm erg \, s^{-1} \,,
\end{equation}
exceeding the star's surface luminosity by several orders of magnitude, and more than a hundred times its Eddington luminosity, suggesting that the star cannot provide the required energy dissipation rate over multiple orbits.    

\subsection{Mass transfer}
\label{sec:masstfer}

Mass-transfer through the L2 point can produce orbital decay in close binary stars (e.g., \citealt{Pejcha2017}).  For partial TDEs, however, this effect is small because of the high mass ratio of the companion relative to the central BH. 

As a mass element $\delta m > 0$ is stripped from the star, it carries along some orbital energy 
\begin{equation}
    E_{\rm \delta m} = f_{\rm E} \frac{\delta m}{m_\star} E_{\rm orb} \,,
\end{equation}
where $f_{\rm E}$ is the ratio of the parcel's specific energy to that of the star. The resulting change in orbital period due to the stripping of mass is then given by
\begin{equation}
    \left. \dot{P} \right|_{\rm MT} = \frac{\Delta P}{P} = \frac{3}{2} \frac{\Delta a}{a} = -\frac{3}{2} (1-f_{\rm E}) \frac{\delta m}{m_\star} \,,
    \label{eq:Pdot_from_dm}
\end{equation}
where we neglected terms of order $\delta m/\MBH$, and any of the aforementioned orbital dissipation mechanisms. Since $f_{\rm E}$ is expected to be very close to 1, mass transfer is not likely to provide the necessary magnitude of observed $\dot{P}$. Given the binary's extreme mass ratio, we expect $|1-f_{\rm E}| \lesssim (m_\star/\MBH)^{1/3} \approx 10^{-2}$, accounting for the energy exchange between the star and the stripped material. Thus, the observed $|\dot{P}| \approx 2\times10^{-3}$ may only be explained if $\delta m/m_\star > 0.1$, implying that the star would have been significantly depleted over the 20 observed flares.    {Moreover, existing simulations of partial TDEs find that when the energy change of the remaining bound stellar core is dominated by mass-loss, not tides, the energy change is such that $\dot P > 0$, i.e., the star gets a kick that increases its energy \citep{Manukian2013,Ryu2020,Cufari2023}.  This is incompatible with the $\dot P$ in ASASSN-14ko.}

{The above estimates focus on the case of a partial TDE.   If the mass transfer is dominated by ablation due to interaction with an ambient disc, the mass lost from the star will carry a specific energy very similar to that of the star, i.e., $f_E \simeq 1$.  The mass-loss itself will thus not change the orbital period significantly (eq. \ref{eq:Pdot_from_dm}).   The interaction between the star, the stripped mass, and the ambient disc will change the orbit via gas drag as described and quantified in \S \ref{sec:star_disc}.}

\section{Formation channel for PNTs} \label{sec:formation}

In the previous sections we concluded that ASASSN-14ko involves a somewhat inflated star (possibly due to tidal heating) interacting with a pre-existing AGN disc, with a pericenter distance $\rp$ that is a few times greater than the star's tidal radius, $\rp \approx 4-5 \; \rtidal$. Here we discuss possible dynamical formation channels to produce the observed system.

Consider a star orbiting the SMBH with semi-major axis $a$ and pericenter distance $\rp \ll a$. The star's angular momentum evolves due to two-body relaxation on a timescale \citep[e.g.,][]{Merritt_2013,Linial_Sari_2023}
\begin{equation}
    \tau_{\rm 2B}^{J} \approx \frac{P(a)}{2\pi \ln{\Lambda} \; N_\star(a)} \pfrac{\MBH}{m_\star}^2 \pfrac{\rp}{a} \,,
\end{equation}
where $N_\star(a)$ is the number of stars of semi-major axis of order $\sim a$ and $\ln \Lambda \approx \ln(\MBH/m_\star)$ is the Coulomb logarithm. Here we assumed that all stars are of similar mass, $m_\star$. If an accretion disc is present around the SMBH and extends beyond $\rp$, the star's orbital energy is dissipated through drag on a timescale (e.g., equation \ref{eq:Pdot_avg_rp})
\begin{equation}
    \tau_{\rm disc}^{E} \approx \frac{3P}{2\dot{P}_{\rm disc}} \approx \frac{m_\star P(a)}{2\pi \Sigmad(\rp) R_\star^2} \pfrac{\rp}{a} \,,
\end{equation}
Evolution rates due to disc interaction and two-body relaxation become comparable (e.g., $\tau_{\rm disc}^E \approx \tau_{\rm 2B}^J$) when
\begin{equation}
    \Sigmad(\rp) R_\star^2 \approx m_\star \pfrac{m_\star}{\MBH} \pfrac{a}{\rh}^{3-\gamma} \ln \Lambda \,,
\end{equation}
where we assumed a stellar density cusp $n_\star(r)\propto r^{-\gamma}$ (e.g., $\gamma = 7/4$ for a Bahcall-Wolf cusp) such that $N_\star(a) \approx (\MBH/m_\star) (a/\rh)^{3-\gamma}$, where $\rh$ is the SMBH's radius of influence (defined here as the radius enclosing a total mass of $2\MBH$). Adopting the M-Sigma relation and $\gamma = 7/4$ we obtain
\begin{equation} \label{eq:2B_disc_transition_critertion}
    \Sigmad(\rp) R_\star^2 \approx 10^{-6}\; {\rm M_\odot} \; m_1 \pfrac{P}{10^5 \, \rm yr}^{5/6} M_{\bullet,7}^{-1.1} \,,
\end{equation}
marking the conditions at which the diffusive angular momentum evolution transitions to circularization via disc interaction. As the two-body scattering rate is dominated at $r\approx \rh$ \citep[e.g.,][]{Stone_2020}, most stars undergoing this process will initially have typical orbital periods of order $2\pi \sqrt{\rh^3 / G\MBH} \approx 1.7\times 10^5 \, M_{\bullet,7}^{0.25} \; \rm yr$.
For $R_\star \approx \rm R_\odot$ and $\dot{m}/\dot{m}_{\rm Edd} \approx 10^{-3} - 10^{-2}$ the above condition (Eq.~\ref{eq:2B_disc_transition_critertion}) is satisfied when
\begin{multline} \label{eq:r_p_disc2B}
    r_{\rm p, disc/2B} \approx \\
    2.6 \; {\rm au} \; R_1^{10/3} m_1^{-5/3} \pfrac{\dot{m}/\dot{m}_{\rm Edd}}{10^{-3}} \alpha_{-2}^{-4/3} M_{\bullet,7}^{3.2} \pfrac{P}{10^5 \, \rm yr}^{-25/18}\,,
\end{multline}
where we assumed $\Sigmad \propto r^{-3/5}$, as in the gas-pressure dominated regime, appropriate at this radius (equation \ref{eq:Sigma_gas_disc}).

As $\rp$ approaches $r_{\rm p, disc/2B}$, star-disc interaction begins to dominate, such that $a$ and $P$ decrease, while $\rp$ remains essentially fixed\footnote{Star disc interaction dissipates both orbital energy and angular momentum, yet, if $\Sigmad(r) / r$ is a decreasing function of $r$, the angular momentum dissipation rate is lower than that of orbital energy.}. As $a$ decreases, $\tau_{\rm disc}^E \propto a^{1/2} \propto P^{1/3}$, namely, the system spends less time at shorter orbital periods. Thus, the orbital evolution time at the transition between two-body relaxation and disc-driven dissipation (i.e., when condition \ref{eq:2B_disc_transition_critertion} is satisfied) is approximately
\begin{multline}
    \tau_{\rm disc}^E \approx \tau_{\rm 2B}^J \approx \\ 1.6\times 10^5 \, {\rm yr} \, \pfrac{P}{10^5 \, \rm yr}^{-17/9} R_1^{10/3} m_1^{-5/3} \alpha_{-2}^{-4/3} \pfrac{\dot{m}/\dot{m}_{\rm Edd}}{10^{-3}} M_{\bullet,7}^{4.5} \,,
\end{multline}
much shorter than the typical AGN phase, such that the disc persists throughout the evolution, without undergoing significant changes.

We conclude that in the presence of an extended disc, steadily accreting at a rate $\dot{m}/\dot{m}_{\rm Edd} \gtrsim 10^{-3}$ (such that $r_{\rm p, disc/2B} \gtrsim \rtidal$), stars may ultimately avoid "regular" tidal disruption or partial tidal disruption on a highly eccentric orbit, as disc interaction circularizes their orbit due to disc interaction when $\rp$ is still several times greater than $\rtidal$. {This process has also been discussed in \cite{Macleod_Lin_2020}, who consider the effects of disc-induced drag on loss-cone dynamics}. While these stars escape their tidal-disruption fate, star-disc interaction ultimately grinds down their outer layers through ablation as discussed in \S \ref{sec:ablation}. The ablation mass-loss timescale relative to their circularization time is given by
\begin{equation}
    \frac{\tau_{\rm abl}}{\tau_{\rm P}} = \frac{m_\star \dot{P}} {\delta m} \approx 2\times 10^3 \; R_1^{-2} \pfrac{h}{R_\star} m_1 M_{\bullet,7}^{-2/3} \pfrac{P}{10^5 \, \rm yr}^{2/3} \,
\end{equation}
where we used equation \ref{eq:dM_ablation}. Hence, at sufficiently long orbital periods, $\tau_{\rm abl} \gg \tau_{\rm P}$ and the stellar mass is effectively unchanged. However, as $P$ approaches the observed periodicity seen in ASASSN-14ko, ablation and orbital decay proceed on comparable timescales. Thus, ASASSN-14ko's current state may represent a natural termination point of the discussed evolution, supported by the fact that the accreted mass implied by the observations is $\delta m_{\rm rad}/m_\star \approx \dot{P} \sim \mathcal{O}(10^{-3})$ (see eq. \ref{eq:dM_rad}). 

{In summary, the combination of angular momentum relaxation and star-disc interaction may produce stars on tight, eccentric orbits, with period, period evolution and ablation-driven mass loss compatible with ASASSN-14ko. We note, however, that this process does not preferentially produce stars with $\rp \approx {\rm few}\times\rtidal$ (as seen in equation \ref{eq:r_p_disc2B}, $r_{\rm p,disc/2B}$ strongly depends on the relevant parameters). Nevertheless, it is possible that tidal heating and/or mass transfer are required to produce luminous flares, making the smaller fraction of systems with $\rp \approx {\rm few}\times\rtidal$ observationally selected in time-domain surveys.}

One important implication of the comparison between $\tau_{\rm abl}$ and $\tau_{\rm P}$ is that systems akin to ASASSN-14ko are not likely to evolve to hour-timescale periods, as collisions with the disc become too destructive at periods of order $10^2 \, \rm d$. Mildly eccentric, stellar-EMRIs with with periods of order hours have been recently suggested as the origin of QPEs \citep[e.g.,][]{Linial_Sari_2023,Krolik_Linial_2022,Lu_Quataert_2023,Linial_Metzger_23}. If this indeed the case, QPE formation is likely to be quenched in the presence of an extended AGN disc, as stars are destroyed (or at least substantially stripped) through ablation, at comparatively long orbital periods while their orbits are still highly eccentric, before evolving to the orbits of interest. We therefore conclude that PNTs similar to ASASSN-14ko, where an extended disc is evident, do not ultimately evolve to produce QPEs, unless the star somehow survives the repeating destructive collisions with the disc. We note that the star-disc interaction described in \cite{Linial_Metzger_23} involves a  low-mass, compact accretion disc which originated from a second star undergoing a TDE in the same galactic nucleus, such that star-disc collisions only commence once the EMRI is on few-hour period, as opposed to the case of an extended AGN disc.

An alternative formation scenario for PNTs involves the Hill's mechanism \citep{Hills_1988}, which results in stars on highly eccentric orbits with periods of order years. {This possibility has been recently considered by \cite{Cufari_2022} as the origin of ASASSN-14ko}. A binary of separation $a_{\rm b}$ is tidally ionized when its center of mass approaches a distance $\rtidalb \approx a_{\rm b} (\MBH/m_b)^{1/3}$ from the SMBH, where $m_b \approx 2 m_\star$ is the total mass of the binary, assuming an equal-mass binary. One of the binary components gets ejected as a hyper-velocity star, while the other star remains bound to the SMBH on an eccentric orbit with $\rp \approx \rtidalb$ and $a \approx \rtidalb (\MBH/m_\star)^{1/3}$. 

The orbital period of the bound star is approximately \citep[e.g.,][]{Kobayashi_2012,Cufari_2022}
\begin{equation}
    P_{\rm Hills} \approx \pi \sqrt{\frac{a_b^3}{Gm_\star}} \pfrac{\MBH}{m_\star}^{1/2} = 180 \; R_1^{3/2} m_1^{-1} M_{\bullet,7}^{1/2} \pfrac{a_b}{R_\star}^{3/2} \; \rm days \,,
    \label{eq:PHills}
\end{equation}
and its pericenter distance
\begin{equation}
    \rp \approx r_{\rm tidal,b} = 1 \, \pfrac{a_{\rm b}}{R_\star} R_1 \, m_1^{-1/3} \, M_{\bullet,7}^{1/3} \; \rm au \,.
\end{equation}

For ASASSN-14ko we previously inferred $R_\star$ of order a few $\rm R_\odot$, and $\rp$ a few times $\rtidal$. We note that the combination of these two facts implies $P_{\rm Hills} \gg P_0$, as both $R_1 \gtrsim 2$ and $a_b/R_\star \gtrsim 2$. Thus, if the system has indeed formed though the Hill's mechanism, its initial orbit was considerably longer than is currently observed. Assuming that the currently inferred $\rp$ reflects the original pericenter distance soon after the binary disruption, the system has evolved to its current state from an initial orbital period of order a decade. {In the scenario suggested by \cite{Cufari_2022}, the currently observed $P$ of ASASSN-14ko is interpreted as $P_{\rm Hills}$, namely, the period of the bound component immediately post binary splitting. The challenge with this interpretation is that the star would then have $\rp \approx \rtidal$, where tidal heating would rapidly destroy it, while the system is seen to survive multiple pericenter passages, and its projected remaining lifetime is roughly $\tau_{\rm P} = P/\dot{P} \approx 150 \, \rm yr$.}

A pericenter distance of $\rp \approx 20 \; \rm au$ corresponds to an initial binary separation of $a_{\rm b} \approx \rp (m_\star/\MBH)^{1/3} \approx 20 \, \rm R_\odot$, whose binary orbital velocity of $\sim 140 \, \rm km \, s^{-1}$ is similar to the dynamical velocity dispersion within the SMBH's radius of influence. Namely, the progenitor binary is close to the hard/soft boundary for binaries in the galactic center. 

Given the Hill's mechanism rate of roughly $10^{-5}-10^{-4} \; \rm yr^{-1}$ in galaxies of comparable mass \citep[e.g.,][]{Linial_Sari_2023}, and given the inferred lifetime of roughly $10^{3} \, \rm yr$, approximately $10^{-2}-10^{-1}$ of AGNs should host a low mass star undergoing orbital decay due to the star-disc interaction. Observationally, these systems may display similar behavior to that of ASASSN-14ko, i.e., months-years long periodic flares, with a coherent period decay.

\section{Discussion and Conclusions} \label{sec:discussion}

{The growing class of periodic nuclear transients plausibly associated with stars in orbit around a massive black hole provides a new window into stellar and gas dynamics in galactic nuclei.  A number of the observed sources show secular evolution of their periods over the course of years to decades (e.g.,  \citealt{Payne_2022,Miniutti_23,Miniutti_23a}).   In this paper we have surveyed a range of possible explanations for this secular evolution.   We have focused on the superb timing measurements of ASASSN-14ko's repeating flares \citep{Payne_2022} but our analysis is more broadly applicable and can be readily applied to other periodic nuclear transients.}  For the specific case of ASASSN-14ko, we have shown that its observed $\dot{P}$ most likely originates from gas drag due to the interaction between a star on an eccentric orbit and an accretion flow onto the central SMBH, where the star's orbit is inclined with respect to the disc plane.
We summarize a few key features of our model and their implications:
\begin{enumerate}
    \item The radius at which star-disc collisions occur varies periodically due to relativistic apsidal precession, implying that the star "samples" the accretion flow at the entire range $\rp < r < r_{\rm max} = a (1+e)$ over the course of a precession cycle. Depending on the density profile of the disc, the orbital decay averaged over a full precession timescale is typically dominated by collisions close to $\rp$. Variations in the collision conditions results in period modulation of the per-orbit orbital decay, which may be imprinted in the timing of the flares as residuals of the quadratic fit to the O-C data, as demonstrated in figure \ref{fig:Numerical_Pdot}. {These variations in general include a sinusoidal component due to apsidal precession and a  $\ddot{P}$; the latter is likely somewhat larger than the former for the parameters of ASASSN-14ko, but both may be detectable.} We therefore encourage further analysis of the flare timing data, and its deviations from constant $\dot{P}$.
    \item In our proposed interpretation, the observed period evolution is directly related to the hydrodynamical drag experienced by the star as it ploughs through the disc. It can therefore be considered a \textit{mechanical} measurement of the flow conditions in the vicinity of the SMBH. Repeating nuclear transients akin to ASASSN-14ko may help constrain the physical properties of AGN accretion discs as well as the stellar dynamical processes taking place in galactic nuclei. The flare timing behavior determines the star's orbit, as well its relativistic precession rate, providing an independent measurement of the SMBH mass (and possibly, over longer timescales, its spin). 
    \item The observed $\dot{P}$ constrains the disc surface density $\Sigmad(\rp)$, setting a lower limit on the disc mass contained within the star's pericenter distance. For ASASSN-14ko, we find that the minimal disc mass is several times greater than the mass of the star, implying that the disc material cannot originate from previously stripped stellar material. Instead, an independent, pre-existing accretion disc, like an AGN disc, is more likely, as is indeed observationally evident in ASASSN-14ko's host galaxy \citep{Tucker2021}.
    \item ASASSN-14ko has been previously identified as a repeating/partial TDE candidate, involving a star undergoing episodic Roche lobe overflow at pericenter passage \citep{Payne_2021,Liu2023}.  Our interpretation is both similar and different to that of previous work. In particular, in order to explain the magnitude of the observed $\dot{P}$, we find   that the star's pericenter distance is at most a few times $\rtidal$, and that the star is somewhat inflated, possibly due to tidal heating, which plays an important role in shaping the stellar structure once $\rp \lesssim 4.6 \rtidal$ (eq. \ref{eq:r_MT}).  On the other hand, it is not clear that tidal forces dominate the stellar mass loss. Given the relatively small  $\rp/\rtidal$ required to explain the $\dot{P}$, the star may be undergoing tidally-driven mass loss through Roche lobe overflow at pericenter. However, given the constraints on the disc properties from the measured $\dot P$, it is likely that ablation caused by interaction with the disc strips sufficient mass per orbit to power the observed flares independent of any tidal mass-stripping (see \S \ref{sec:ablation}).  It is also unclear whether the flare radiation in ASASSN-14ko is powered by accretion or by circularization shocks.   The radiative efficiency of the two processes differ by at least a factor of $\sim 10$ and so if circularization shocks power the observed radiation the integrated stellar mass-loss to date to explain the observed flares is $\gtrsim 0.2 M_\odot$ (eq. \ref{eq:dM_rad}); the system is thus likely to undergo significant secular evolution over the coming few decades.  In our view, the rough constancy of the flares to date and the presence of strong X-ray radiation somewhat favor accretion powering the flares.  However, the short viscous time required and the relative timing of the X-ray and optical lightcurves \citep{Payne_2022} are not easy to understand.
    \item For PNTs in which ablation due to star-disc interaction dominates the mass-loss, apsidal precession changes the time between star-disc collisions and could lead to the appearance of two flares per orbit during the phase of the precession cycle when the two star-disc collisions are particularly well-separated in time (eq. \ref{eq:2flares}).
    \item{{Our interpretation and modeling of ASASSN-14ko has several broader implications for theoretical modeling of partial TDEs and periodic nuclear transients.   First, the cumulative impact of tidal heating on the structure of the star must be accounted for in determining the mass transfer and thus the properties of the observed flares.  Tidal heating almost certainly leads to mass transfer occurring at pericenter distances a factor of few larger than the usual $r_p \simeq 2 \rtidal$ condition inferred in single passage tidal disruption calculations (eq. \ref{eq:r_MT}).   The structure of the star at that point will be significantly different from a standard (non-tidally heated) stellar model, which presumably will change the amount of mass stripped per orbit relative to calculations that do not include the cumulative impact of tidal heating (e.g., \citealt{Ryu2020,Liu2023}).  Second, if star-disc interactions are also important, as in our model of ASASSN-14ko, the stellar structure is further strongly modified by the repeated shocks generated as the star collides with the disc.}}
    \item Star-disc interactions have been invoked to explain the origin of QPEs \citep{Xian_2021,Linial_Metzger_23, Franchini_23}. Unlike the situation described in this paper, \cite{Linial_Metzger_23} considered a main-sequence star on a mildly eccentric ($e\approx0.1$) orbit, impacting a compact ($r\approx {\rm few}\times \rtidal$) accretion disc twice per orbit, with an orbital period of order hours. Disc drag is expected to dissipate the QPE period at a rate
    \begin{equation}
        \dot{P}_{\rm QPE} \approx -2\times 10^{-5} \; \pfrac{P_{\rm QPE}}{2 \rm \, hr} \alpha_{-2}^{-1} M_{\bullet,6}^{-1} \pfrac{\dot{m}/\dot{m}_{\rm Edd}}{10^{-2}}^{-1} m_1^{-1} R_1^2 \,,
    \end{equation}
    where $P_{\rm QPE} \approx P/2$ is the average QPE recurrence time. Here we considered a nearly circular orbit, $\rp \approx a$ and assumed a radiation-pressure dominated disc (eq. \ref{eq:Sigma_rad_disc}), as is appropriate for the periods of interest. These predictions concerning the period evolution of QPEs could be compared against long-term monitoring of their flare recurrence pattern and provide an additional observational test to theoretical models. 
    {While star-disc interaction appears to be the most promising explanation in the case of ASASSN-14ko, it might be inadequate to explain the temporal behavior of the repeating TDE candidate eRASSt J045650.3 \citep{Liu_obs_2023}, which has been observed to flare repeatedly roughly once every $\sim 223 \, \rm d$. Recent observations (Liu, private communication) suggest a period decay at a rate of $\dot{P} \sim \mathcal{O}(0.1)$, greatly exceeding the maximal drag-induced orbital decay (obtained when $\Sigmad = \Sigma_{\rm disc,max}$, eq. \ref{eq:Sigmad_max}), of order $\dot{P}_{\rm drag,max} \approx 10^{-3} \, (P_{\rm orb}/223 \,{\rm d})^{13/24} \, R_1^{19/16} \, m_1^{-35/48} \, \alpha_{-2}^{-7/8}$, suggesting that an alternative mechanism is at play.}
    \item We discussed the ablation of stellar envelopes caused by star-disc interaction in \S \ref{sec:ablation}. If a significant fraction of the ablated material remains in the disc, this process may pollute the inner regions of the disc with stellar material. More broadly, ablation and/or tidal stripping of stars could play an important role in enhancing AGN metallicity, which has been observed to exceed several times the solar metallicity \citep[e.g.,][]{Hamann_99,Juarez_09}. 
\end{enumerate}

\section*{Acknowledgements}   We thank Wenbin Lu, Riccardo Arcodia, Brian Metzger, Anna Payne, Ben Shappee and Barry McKernan for valuable conversations, and the referee for valuable suggestions.  IL acknowledges support from a Rothschild Fellowship and The Gruber Foundation. This work was supported in part by a Simons Investigator grant from the Simons Foundation (EQ).   This research benefited from interactions at workshops funded by the Gordon and Betty Moore Foundation through grant GBMF5076.

\section*{Data Availability}
The data underlying this article will be shared on reasonable request to the corresponding authors.



\bibliographystyle{mnras}
\bibliography{main} 



\appendix

\section{Evolution due to tides}
\label{appendix:tides}
\subsection{Tidal Energy Input} 
For a single parabolic orbit, the energy deposition in the stellar envelope per orbit is given by \citep[e.g.,][]{Press_Teukolsky_1977,Lee_Ostriker_1986}
\begin{equation}
\label{eq:Etides}
    E_{\rm tides} \approx E_\star \pfrac{\rp}{\rtidal}^{-6} T_2(\rp/\rtidal) \,,
\end{equation}
where $T_2$ is a dimensionless coupling coefficient. The star's structure will be significantly modified if the average tidal heating rate, $\dot{E}_{\rm tides} = E_{\rm tides}/P$ exceeds the star's surface luminosity, $L_\star$, corresponding to
\begin{equation}
\label{eq:rptapp}
    \rp \gtrsim \rtidal \pfrac{\tau_{\rm KH}}{P}^{0.08} \approx 4.2 \; \rtidal \pfrac{P}{1 \, \rm yr}^{-0.08} \,,
\end{equation}
where $\tau_{\rm KH} = E_\star/L_\star$ is the star's Kelvin-Helmholtz timescale. Here we approximated $T_2(\rp/\rtidal) \approx (\rp/\rtidal)^{-6}$ based on Figure 1 of \citet{Lee_Ostriker_1986} for an n = 3 polytrope (appropriate for a solar-type star); $T_2$ falls off even more rapidly with increasing $\rp/\rtidal$ for fully convective stars (n = 3/2 polytropes in \citealt{Lee_Ostriker_1986}).   Equation \ref{eq:rptapp} is equivalent to equation \ref{eq:r_MT} of the main text.

Using equation \ref{eq:tnlsolar} from the next section, we estimate that at the onset of significant structural changes due to tidal heating (and likely mass transfer), i.e., when equation \ref{eq:rptapp} is satisfied, the nonlinear damping time of the tides with energy given by equation \ref{eq:Etides}, satisfies $t_{nl}/P \simeq 0.3 (P/{\rm yr})^{-3/2} (L_\star/L_\odot)^{-1/2}$ (this is for solar type stars).   For orbital periods $\gtrsim 0.5$ yr, the nonlinear damping time is short compared to the orbital period and so the single passage estimate of the tidal energy in equation \ref{eq:Etides} is reasonable.  For shorter orbital periods, however, the nonlinear damping time is longer and the tidal energy may accumulate over multiple orbits.

\subsection{Non-linear Tidal Dissipation} 

For a bound star on a highly eccentric orbit, equation \ref{eq:Etides} remains appropriate so long as the timescale to dissipate the tidal energy is shorter than the orbital period, so that the tidal energy is dissipated prior to the next pericenter passage.   For the tidal energies relevant to partial TDEs or near-grazing encounters with $\rp \sim \rtidal$, the dominant tidal dissipation processes are almost certainly a combination of nonlinear coupling to other stellar modes (e.g., \citealt{Kumar, Weinberg2012}), wave breaking, and/or shocks.   The most detailed study relevant to the present work is that of \citet{Kumar} who estimate non-linear damping times as a function of tidal energy for solar type stars and fully convective stars:
\begin{equation}
t_{nl} \simeq 1 \, {\rm hr} \, \left(\frac{E_{\rm tides}}{10^{48} \, {\rm erg}}\right)^{-1/2} \ \ \ {\rm (solar-type \, stars)}
\label{eq:tnlsolar}
\end{equation}
\begin{equation}
t_{nl} \simeq 1 \, {\rm hr} \, \left(\frac{E_{\rm tides}}{10^{48} \, {\rm erg}}\right)^{-1} \ \ \ {\rm (convective \, stars)}
\label{eq:tnlconv}
\end{equation}
The different scalings with tidal energy in equations \ref{eq:tnlsolar} and \ref{eq:tnlconv} are because the nonlinear coupling is via resonant excitation of internal gravity waves in the solar-type star case but non-resonant excitation of sound waves in the fully convective star case.   

In equations \ref{eq:tnlsolar} and \ref{eq:tnlconv}, we have scaled the tidal energy to a value similar to that needed to explain the orbital evolution of ASASSN-14ko via tidal energy dissipation (eq. \ref{eq:dEorb_obs}).   We stress, however, that at  such large tidal energies (of order the stellar binding energy), the nonlinear perturbation theory calculation of \citet{Kumar} is very uncertain.   Nonetheless, the nonlinear dissipation timescale is extraordinarily short, implying that equation \ref{eq:Etides} is a reasonable estimate and that the tidal energy is rapidly transferred to the star during/just after pericenter.  As argued in \S \ref{sec:tides} this makes it very difficult to understand the orbital evolution of ASASSN-14ko as due to tides, since the star would have had to survive a cumulative tidal energy dissipation significantly larger than the stellar binding energy over the observed $\sim$ 20 flares.   This issue is particularly acute in solar type stars where the coupling to internal gravity waves occurs in the stellar interior \citep{Kumar, Weinberg2012} and thus to the bulk of the stellar mass.   The only way out of this conclusion appears to be if the dominant nonlinear damping mechanism is not coupling to internal stellar modes, but rather shocks near the stellar surface, so that most of the tidal energy is transferred to just a small fraction of the mass each orbit.   Numerical experiments focused on this question would be valuable for understanding the evolution of stars close to massive BHs in galactic nuclei.  

\section{Gravitational focusing cross section} \label{appendix:graviational_focusing}
The regime $R_{\rm drag} \approx Gm_\star/v_{\rm rel}^2$ is appropriate for compact companions with $R_\star < \rp \frac{m_\star}{\MBH}$. For such objects, tidal gravity is never important, given that
\begin{equation}
    \frac{\rp}{\rtidal} \gtrsim \pfrac{\MBH}{m_\star}^{2/3} \gg 1 \,.
\end{equation}

In this regime, the disc surface density implied by the observations is approximately
\begin{multline}
    \Sigmad \approx \frac{\MBH}{3\pi a^2} \pfrac{a}{\rp} \pfrac{\MBH}{m_\star} \dot{P} \approx \\
    2.5 \times 10^6 \; M_{\bullet,7}^{1/3} \mathcal{P}_{115}^{-4/3} \pfrac{\dot{P}}{2.6\times 10^{-3}} \pfrac{a}{\rp} \pfrac{\MBH}{m_\star} \; \rm g \, cm^{-2} \,,
\end{multline}
where generally $\MBH \gg m_\star$ and $a \gg \rp$, so the necessary surface density is several order of magnitude greater than the geometrical cross section regime (equation \ref{eq:Sigmad_from_Pdot}). The corresponding disc mass is given by
\begin{multline}
    m_{\rm d} \approx \pi \rp^2 \Sigmad \approx \frac{1}{3} \MBH \pfrac{\MBH}{m_\star} \pfrac{\rp}{a} \dot{P} \approx \\
    8.7\times10^3 \; \pfrac{\dot{P}}{2.6\times 10^{-3}} \pfrac{\MBH}{m_\star} \pfrac{\rp}{a} \; \rm M_\odot \,.
\end{multline}

Considering that $\rp > R_{\rm ISCO}$, we find a minimal disc mass of
\begin{equation}
    m_{\rm d} > 50 \pfrac{\MBH}{m_\star} \; \pfrac{\dot{P}}{2.6\times 10^{-3}} M_{\bullet,7}^{5/3} \mathcal{P}_{115}^{-2/3} \; \rm M_\odot \,.
\end{equation}

The need for extremely high surface densities and such a massive disc, containing many solar masses, renders this scenario unlikely. Even if the secondary is an IMBH with $m_\star \approx 10^5 \; \rm M_\odot$, one would require an unrealistically massive accretion disc to explain the observed period evolution.


\bsp	
\label{lastpage}
\end{document}